\documentclass[final]{ias2}

\usepackage{graphicx} 
\usepackage{multirow}
\usepackage{array} 

\usepackage{hyperref} 
\newcommand{\pt}{$p_{\rm{t}}$~}

\newcommand{\nudyn}{$\nu_{(+-,dyn.)}$}

\begin{document}

\markboth{Heavy Ions: Results from the LHC}{Tapan K. Nayak}

\title{Heavy Ions: Results from the Large Hadron Collider}

\author[sin]{Tapan K. Nayak} 
\email{Tapan.Nayak@cern.ch}
\address[sin]{Variable Energy Cyclotron Centre, Kolkata - 700064,
  India}

\begin{abstract}

On November 8, 2010 the Large Hadron Collider (LHC) at CERN collided first
stable beams of heavy ions (Pb on Pb) at center-of-mass energy of
2.76 TeV/nucleon. The LHC worked exceedingly well during its one month
of operation with heavy ions, delivering about $10\mu$b$^{-1}$ of
data, with peak luminosity reaching to
 $L_{0} = 2 \times 10^{25}{\rm cm}^{-2}{\rm s}^{-1}$ towards the end of the run. 
Three experiments, ALICE, ATLAS and CMS, 
%The ALICE experiment 
%which was preparing for this data for last
%twenty years, as well as the other two large experiments ATLAS and CMS
recorded their first heavy ion data, which were analyzed in a record  time.
The results of the multiplicity, flow, fluctuations, and
Bose-Einstein correlations indicate that the fireball formed in
nuclear collisions at the LHC is hotter, lives longer, and expands to
a larger size at freeze-out as compared to lower energies.
We give an overview of these as well as new results on 
quarkonia and heavy flavour suppression, and jet energy loss.

\end{abstract}
\maketitle

% \tableofcontents
% \listoffigures
% \listoftables

\section{Introduction}

The Large hadron Collider (LHC) at CERN is 
the world's largest and highest energy accelerator, designed to
address some of the most fundamental questions of recent times
such as, whether Higgs particle 
%(responsible for giving mass to all particles) 
exist or not, why laws of physics are not 
symmetrical
between matter and anti-matter, what makes up the mysterious dark
matter, and how do we understand the primordial state of matter after the Big
Bang and before the formation of nucleons. The answers to each of
these questions may not come that easy and may open up newer
avenues in our quest to discover the mysteries of our Universe. 
The approach at LHC has been to address these issues by studying the
particles produced in
proton-proton \mbox(p-p) as well as ion-ion (such as \mbox{Pb-Pb}) 
collisions at extremely high energies. 
%The physics process through which \mbox{Pb-Pb} collision proceeds and
%the matter produced may be very different from those of the \mbox{p-p}
%collisions. 
With the heavy ion collisions, the aim is to study the primordial
state of matter consisting of quarks and gluons and verify the
predictions of the Standard Model in conjunction with quantum
chromodynamics (QCD).

%Quarks and gluons behave like free particles when they are very 
%close together, but feel much stronger forces when separated. 
%This feature of QCD known as asymptotic freedom, states that
%interaction between quarks becomes very weak at short distances, and
%as a result they behave like free particles.
%Two most novel features of QCD are: (i) asymptotic freedom, which 
%states that quarks behave almost like free particles when they are
%close together, but their interaction increases as they get
%separated, and (ii) quark confinement, suggesting that free quarks are
%never seen in isolation. 

Asymptotic freedom is a feature of QCD which states that 
at short distances the interaction between quarks becomes very weak,
and as a result they behave like free particles.
As we try to increase the distance between pairs of quarks, 
they will have more affinity for each other and hence will be
interacting even more strongly with each other. So, it is almost
impossible to isolate a free quark. 
It has been postulated that the early Universe, immediately after the
Big Bang, consisted of asymptotically free quarks and gluons.
Statistical QCD calculations predict that at high temperature and/or
energy density a system of strongly interacting particles, consisting
of quarks and gluons, is formed where the particles would interact
fairly weakly due to asymptotic freedom.
Such a phase consisting of (almost) free quarks and
gluons is termed as the quark gluon plasma (QGP)~\cite{satz,kapusta}.
By colliding two relativistically accelerated heavy ions, 
it is possible to compress and heat the nuclei to such an extent that
their individual protons and neutrons overlap, creating a region of enormously 
high energy density, where a relatively large number of free quarks 
and gluons can exist for a brief time.
The experimental search for QGP began in the eightees with several
fixed target experiments at Brookhaven National Laboratory (BNL) and
CERN. The results from the experiments at Super Proton Synchrotron (SPS)
led CERN to announce the evidence for a new state of matter in the
year 2000. In the same year, the Relativistic Heavy Ion Collider
(RHIC), a dedicated machine for QGP search became operational at BNL.
The results of last decade of data taking at RHIC points to the creation of a
new form of matter that behaves like a strongly interacting near perfect liquid.
LHC, with an increase of energy by more than an order of magnitude
compared to RHIC,
provides an enormous opportunity to explore the new state of matter in great detail.

At the LHC, the first successful proton-proton (\mbox{p-p}) collisions took place in
2009, and a year later in November 2010, first collisions of
\mbox{Pb-Pb} ions were established at $\sqrt{s_{NN}}= 2.76$~TeV.
%In the span of a month, LHC delivered close to
%10$\mu$b$^{-1}$ of data, which were analyzed in a record time to provide some of
%the most amazing findings so far. 
In this report, we discuss the
capability of LHC experiments for heavy ion collisions, 
and make a short review of the 
the recent results. Section~2 gives a brief introduction to heavy-ion
collisions, and in section~3 we discuss about the experiments. In
section~4, we discuss freeze-out conditions and global properties in
terms of multiplicity and pseudorapidity distributions, 
size of the fireball, elliptic and radial flow, and fluctuations and
correlations. The results on quarkonia and heavy
flavour are given in section~5, section~6 discusses electroweak probes
and in section~7 we give the hard probes of hot and dense matter.
We conclude with a status summary and a future outlook.

\section{Probing the hot and dense matter}

Knowledge of the space-time evolution of the system produced in high
energy heavy ion collisions provides insight into the dynamics of
nuclear matter under extreme conditions. Schematic of the time 
evolution in case of collision of two Lorentz contracted nuclei at
very high energy is shown in Fig.~\ref{fig:evolution}. 
As the colliding nuclei recede from each other, a large amount of energy is deposited in a small region of
space and in a short duration of time. The matter thus created may
have very high energy density and temperature, sufficient to form a 
baryon free region of QGP. The hot and dense plasma 
may not be initially in thermal equilibrium. Subsequently thermal
equilibrium might set in after which the evolution may be governed 
by the laws of thermodynamics. As the plasma expands and cools,
hadronization takes place and after some time the interactions sieze
(freeze-out). Various stages during the collision can be probed by different
observables, such as:

\begin{figure}[ht]
\begin{center}
\includegraphics[width=0.7\columnwidth]{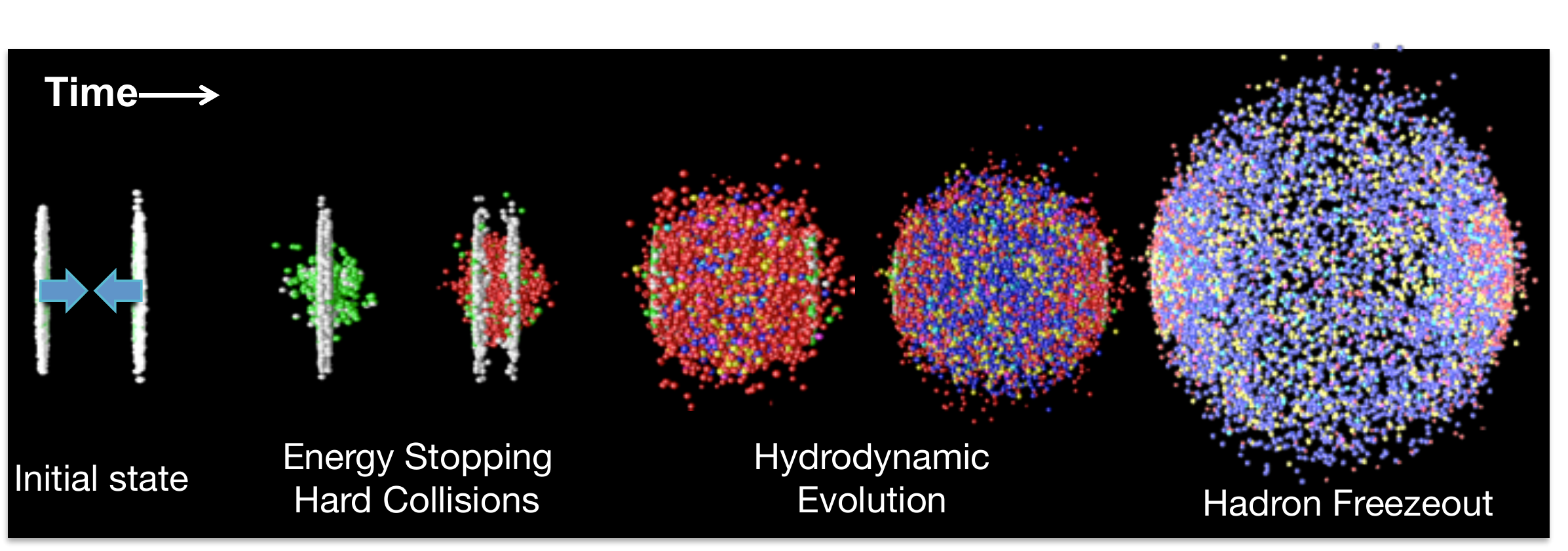}
\caption{
The time evolution of a high energy heavy ion collision. 
}
\label{fig:evolution}
\end{center}
\end{figure}

\begin{itemize}
\item{\bf Freeze-out conditions and global properties:}
Majority of the produced particles in the high energy nuclear
collisions are emitted at freeze-out. A detailed study of these
particles is crucial to estimate the energy density, pressure,
entropy ($s$) as a function of the temperature ($T$) and baryon 
chemical potential ($\mu_B$).  These quantities may be derived from
measurements of multiplicity and rapidity 
rapidity distributions, \pt (transverse momentum)
distributions and transverse energy density 
(dE$_{\rm T}$/dy).
Particle flow and azimuthal
asymmetry, size of the fireball from intensity interferometry, 
correlations and fluctuations, resonance production, particle ratios,
etc. provide crucial information about the 
thermodynamics of the system. 

\item{\bf Electromagnetic probes:}
The electromagnetic probes, i.e. photons and dileptons 
are produced at different phases of evolution.
They are not distorted by final state interactions and once 
produced can escape the interaction region, unaffected,
carrying to the detectors information about the
conditions and properties of the medium at the time of their production.
The measurement of direct thermal radiation can be used 
as a direct fingerprint of the hot and dense medium.

\item{\bf Quarkonia and heavy flavour:}
Quarkonia ($J/\psi$, $\psi'$, $\chi_c$ and $\Upsilon$ family) 
production is considered to provide an unique signature of QGP~\cite{satz}.
It is a sensitive probe of the hot and dense matter and
of the gluon distributions and their modifications in nuclei. 
The suppression of $J/\psi$ production 
has long been predicted as an important probe of a QGP 
formation~\cite{satz}, which
occurs because a c$\bar{c}$ pair formed by fusion
of two gluons from the colliding nuclei cannot effectively bind inside the QGP
because of Debye screening.
Excited states of the c$\bar{c}$ system, such as $\psi'$ are more
easily dissociated and should be largely suppressed.
For the heavier $\Upsilon$($b\bar{b}$) shorter screening lengths 
are required than for the charmonium states. 
Heavy flavor production, open and hidden, is
considered among the most important probes for study of QCD 
properties of the QGP. 

\item{\bf Electroweak probes:}
With the increase of center-of-mass energy at the LHC, electroweak
boson measurements are possible for the first time in heavy ion
collisions. 
The lifetimes of $W$ and $Z$ bosons are quite
short and they decay within the medium, and go unaffected through the
hot and dense matter. Since leptons lose negligible energy 
in the medium, be it partonic or hadronic, the leptonic decay channels
of $W$ and $Z$ may provide information about the initial state in
heavy ion collisions.

\item{\bf Hard probes:}
%In high energy heavy ion collisions, the presence of a large energy 
%scale implies that a variety of distances can be accessed, from the
%small distances
%of the very short times to the long distances at later times. 
The availability of large amount of energy in the very early part of
the collision gives rise to a subset of high 
transverse momentum processes which take place independent of the bulk, 
with the outgoing partons subsequently propagating through the bulk
medium. Jet quenching and energy loss of high \pt hadrons
constitute the most important hard probes, which play important role in 
determining the properties of hot and dense QCD matter.
\end{itemize}

\section{LHC and its experiments}

The accelerator complex at CERN is a succession of particle accelerators
that can reach increasingly higher energies, starting with the
duoplasmatron source for the protons and ECR ion source for the ions,
and ending up at the LHC.  The ions from the source are injected
into a linear accelerator (LINAC3) and to the Low Energy Ion Ring
(LEIR). After that the ions pass through the PS booster, then the Proton
Synchrotron (PS), followed by the Super Proton Synchrotron (SPS) 
and finally to the LHC. For \mbox{Pb-Pb} collisions, a peak luminosity of 
$L_{0}=10^{27}{\rm cm}^{-2}{\rm s}^{-1}$ 
at $\sqrt{s_{NN}}= 5.5$~TeV
has been kept as a final goal. For the first year, LHC delivered
\mbox{Pb-Pb} collisions at $\sqrt{s_{NN}}= 2.76$~TeV with peak
luminosity reaching to $L_{0} = 2 \times 10^{25}{\rm
cm}^{-2}{\rm s}^{-1}$.
In addition to \mbox{Pb-Pb}, LHC is capable of providing collisions of
other species such as \mbox{Ar-Ar} and asymmetric systems of \mbox{p-Pb}.

%This resulted in a
%collection of 30~M minimum bias interactions for the ALICE experiment
%and similar numbers for the other two experiments.
%Next we discuss selected results from the data taken in 2010.

%\begin{figure}[ht]
%\begin{center}
%\includegraphics[width=0.7\columnwidth]{./graph/LHC_ring.pdf}
%\caption{
%The path of the ions in the CERN accelerator complex. The ions pass through the LIE%R, then to the PS, the SPS and finally to the LHC.
%}
%\label{fig:ring}
%\end{center}
%\end{figure} 

%Collisions at the LHC take place at four intersection points, which
%house six experiments, ATLAS, CMS, ALICE, LHCb, TOTEM and LHCf.

\begin{figure}[t]
\begin{center}
\includegraphics[width=0.52\columnwidth]{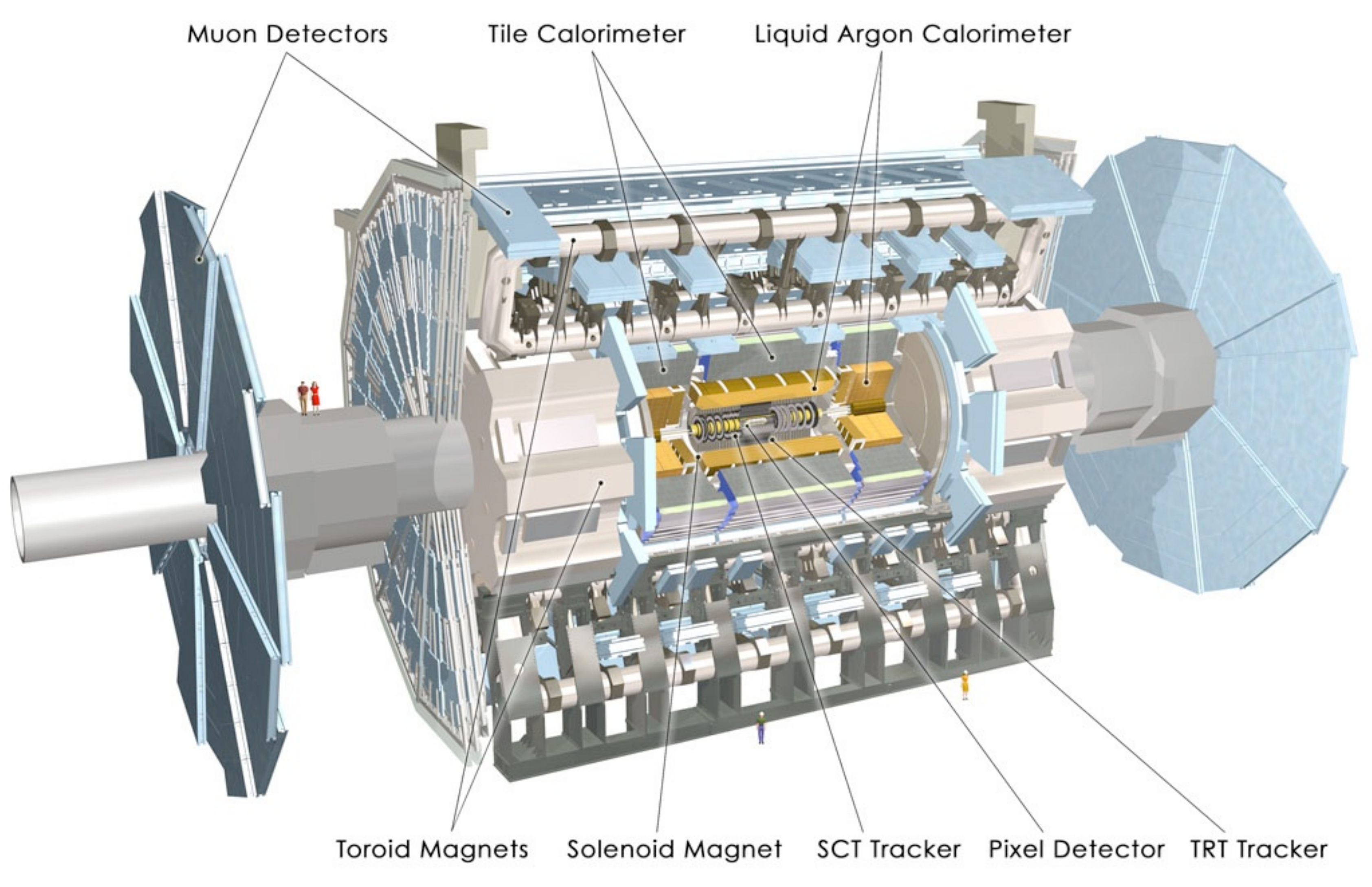}
\includegraphics[width=0.47\columnwidth]{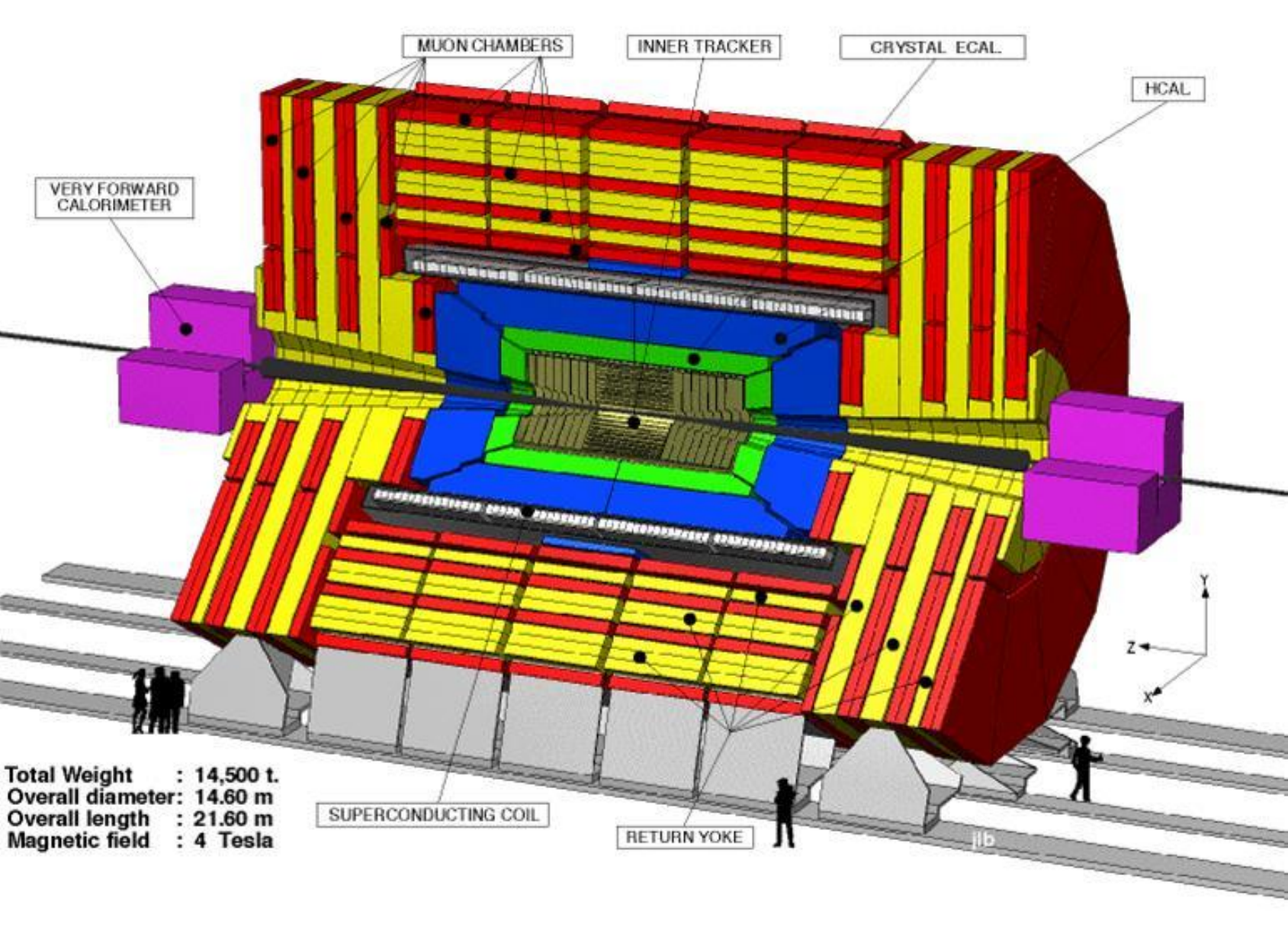}
\includegraphics[width=0.49\columnwidth]{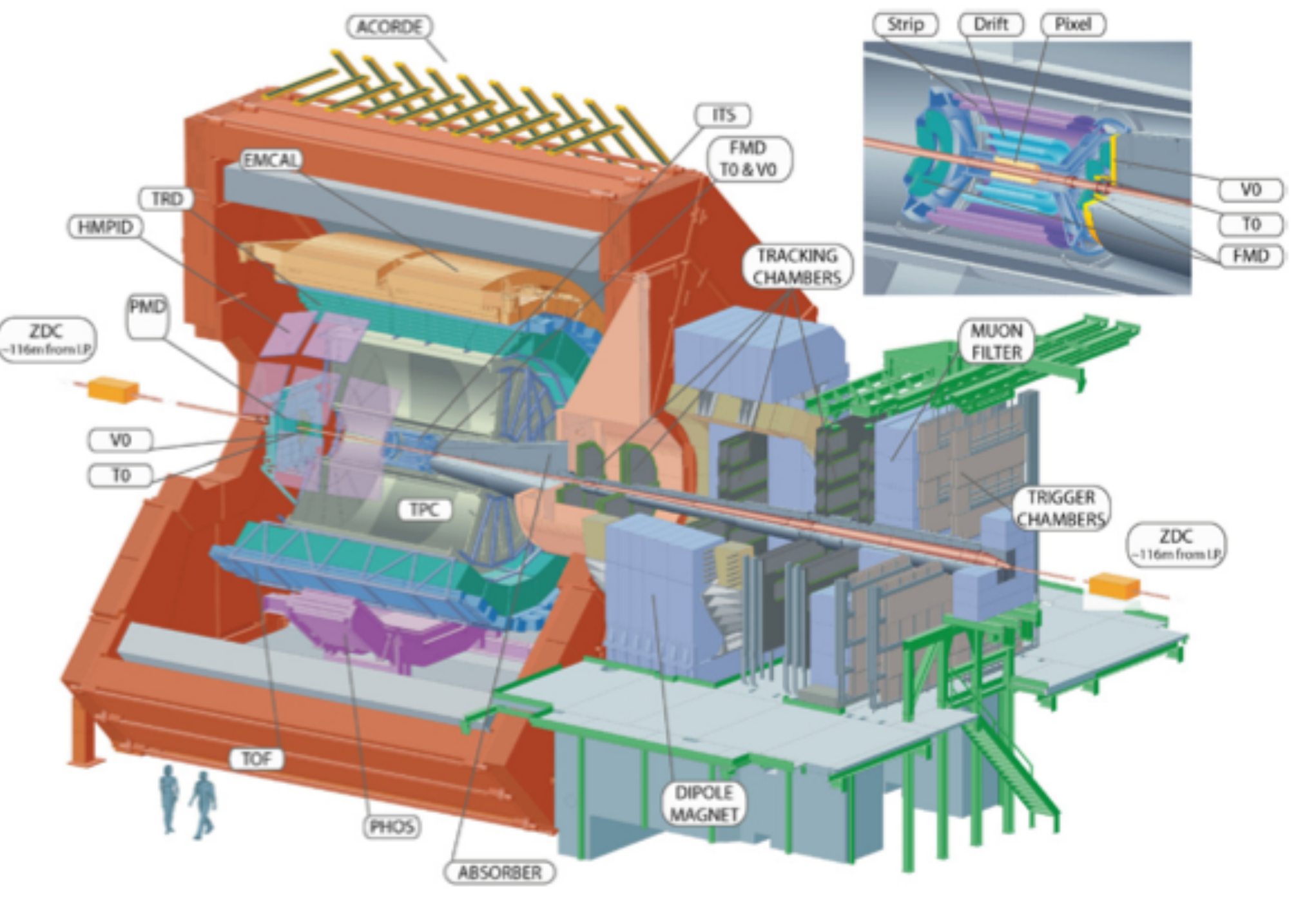}
\includegraphics[width=0.4\columnwidth]{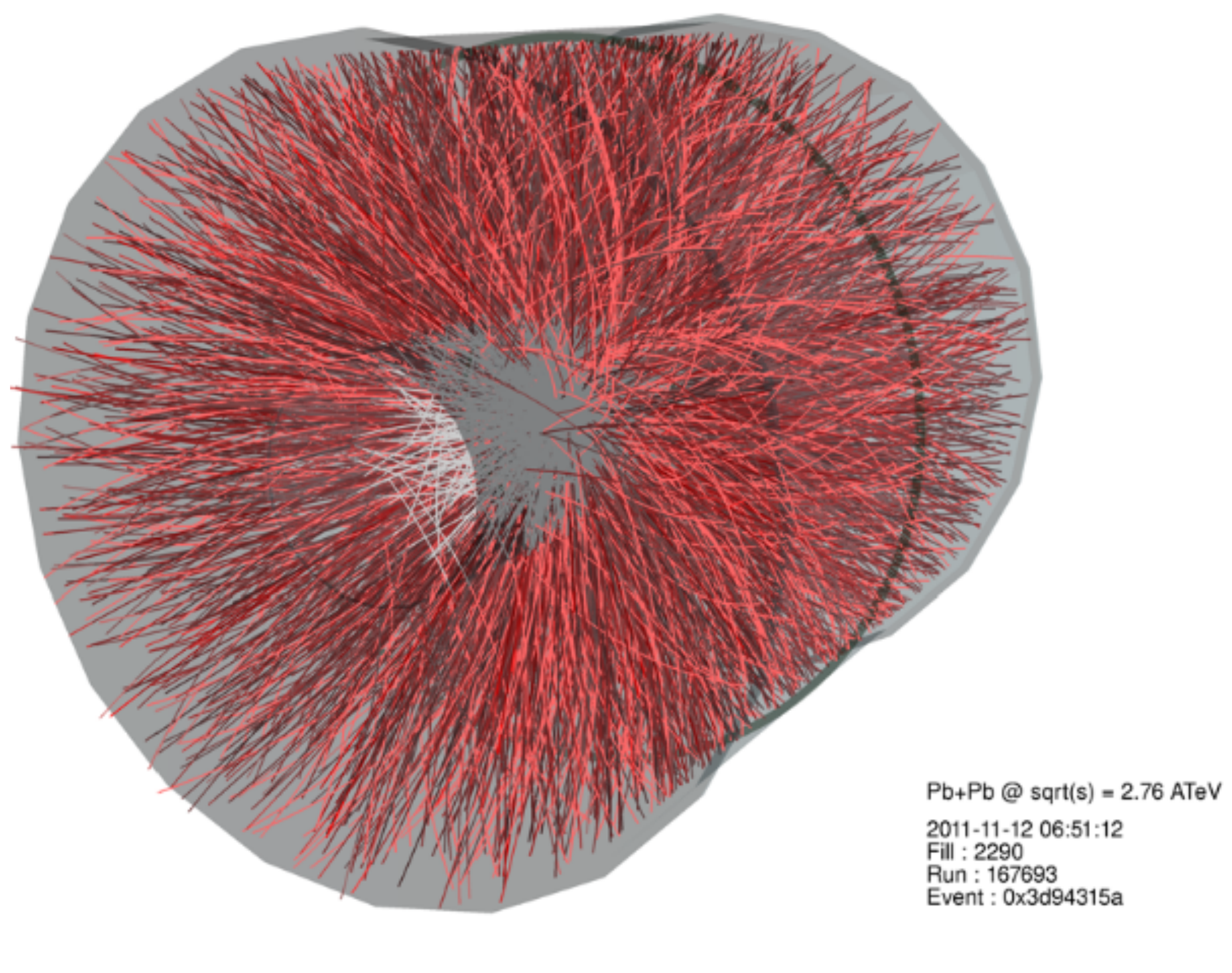}
\caption{Schematic diagram of the ATLAS (top-left), CMS (top-right) and
ALICE (bottom-left) detectors, and an event display (bottom-right) for the
\mbox{Pb-Pb} collisions at $\sqrt{s_{NN}}= 2.76$~TeV in ALICE.}
\label{fig:expt}
\end{center}
\end{figure} 

Sketches of the three experiments, ALICE, ATLAS and CMS
are shown in Fig.~\ref{fig:expt}. An event display 
of charge particle tracks in the ALICE experiment 
obtained for one of the \mbox{Pb-Pb} events is also shown in the figure.

The ALICE experiment~\cite{ALICE} is specifically designed for heavy ion
collisions. The low material budget and low magnetic ﬁeld of the central barrel
makes the experiment sensitive to low-\pt particles. The
central barrel provides powerful tracking with excellent momentum
resolutions, particle identification and capability for measuring
high energy jets. The forward region is quipped with a muon arm,
forward charged particle multiplicity detector(FMD) and photon multiplicity
detector (PMD). The zero degree calorimeter and a set of scintillator
detectors (Vzero, T0 and Acorde) provide timing and used in trigger.

Both the ATLAS~\cite{ATLAS} and CMS~\cite{CMS} detectors incorporate a broad suite of 
high precision subsystems which were originally optimized for very 
high energy \mbox{p-p} collisions,
but also provide capabilities for studying nuclear collisions. 
The large acceptances of both the experiments allow for detailed
study of several observables, and in particular high \pt jets.

\section{Freeze-out conditions and global properties}

\begin{figure}[ht]
\begin{center}
\includegraphics[width=0.9\columnwidth,height=5cm]{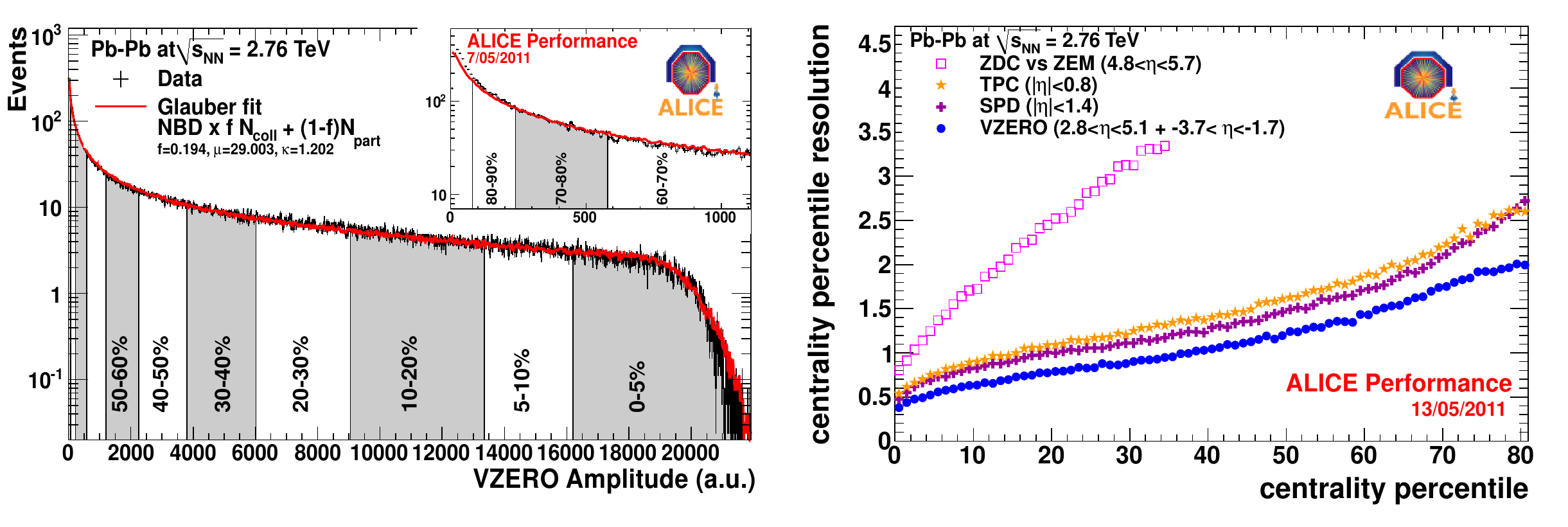}
\caption{Left panel shows the centrality selection in ALICE and the
  right panel gives the centrality
resolution for all centralities~\cite{ALICEcharged}.
}
\label{fig:centrality}
\end{center}
\end{figure}

To start with, we first discuss centrality selection in heavy ion collisions.
Freeze-out conditions, global properties and many other quantities
are best expressed in terms of collision
centrality, which is the overlap of the two colliding nuclei.
Theoretically, the collision centrality is defined
in terms of impact parameter (which is the perpendicular distance between the
centers of the two colliding nuclei) and essentially  indicates the number
of participating nucleons ($N_{\rm part}$). Experimentally, several measured quantities 
scale with the centrality, and may be used as centrality estimators. For
example, in the ALICE experiment the centrality is expressed in terms
of the distribution of VZERO amplitude as shown in the left panel of Fig.~\ref{fig:centrality}.
The peak on the left corresponds to most peripheral collisions, 
the plateau to the mid-central and the edge to the central
collisions. 
The centrality resolution is an indicator for how well the
centrality is known. Fig.~\ref{fig:centrality} (right) gives the
resolutions for different centrality estimators. The centrality 
resolution ranges from 0.5\% in central to 2\% for peripheral
collisions. 

The collision centrality in ATLAS is estimated using the total
transverse energy measured in the Forward Calorimeter (FCal, 
covering 3.2$<|\eta|<$4.9. The collision centrality in CMS is 
determined using the total sum of transverse energy in reconstructed
towers from both positive and negative fiber calorimeters (HF)
covering 2.9$ < |\eta| <$ 5.2. 
In many cases, centrality for specific event classes may also be expressed as a percentage of
the inelastic nucleus-nucleus interaction cross section.
Below we discuss some of the global observables from the first LHC
heavy ion data.

\subsection{Particle multiplicities}

\begin{figure}[t]
\begin{center}
\includegraphics[width=0.47\columnwidth,height=5.2cm]{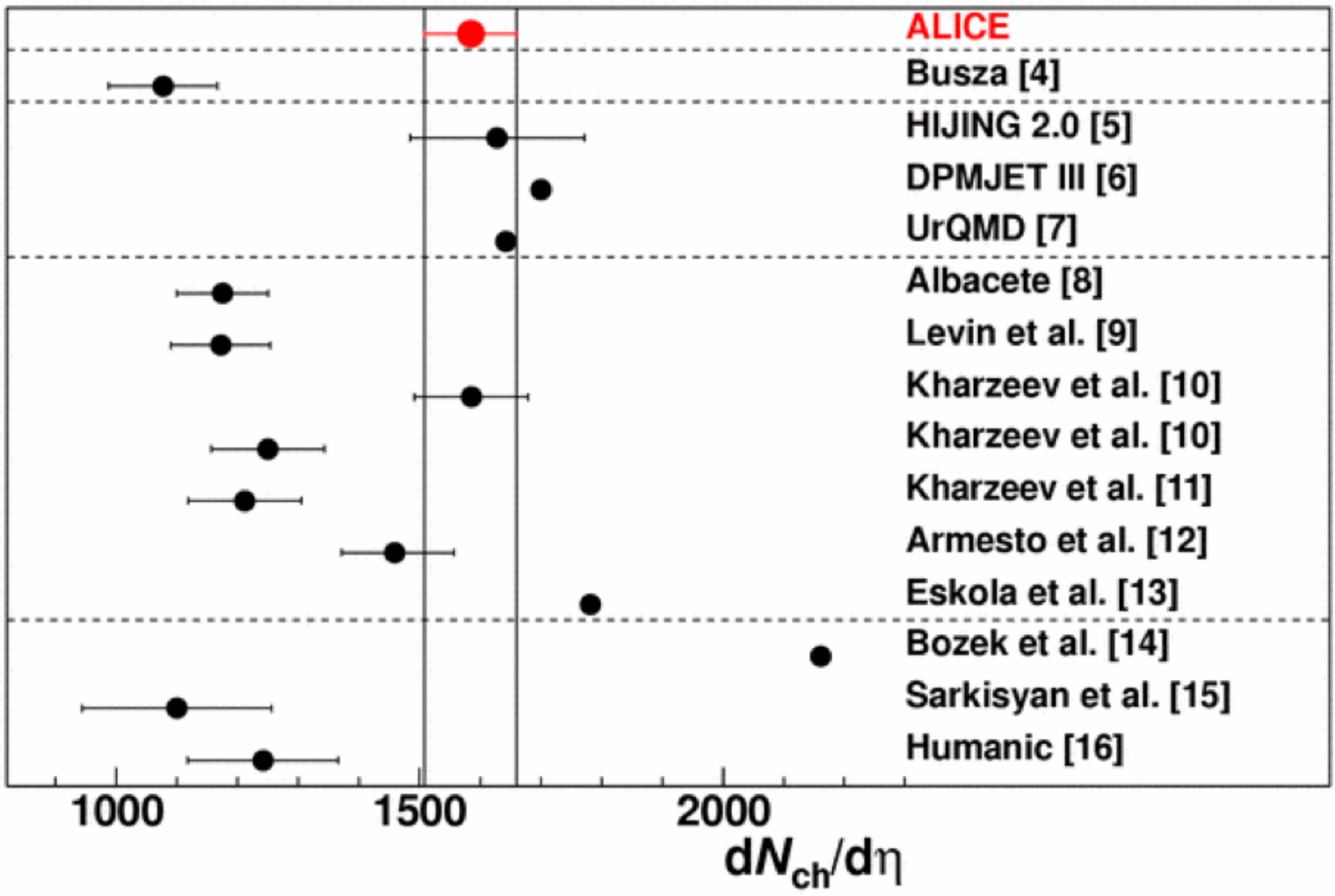}
\includegraphics[width=0.47\columnwidth]{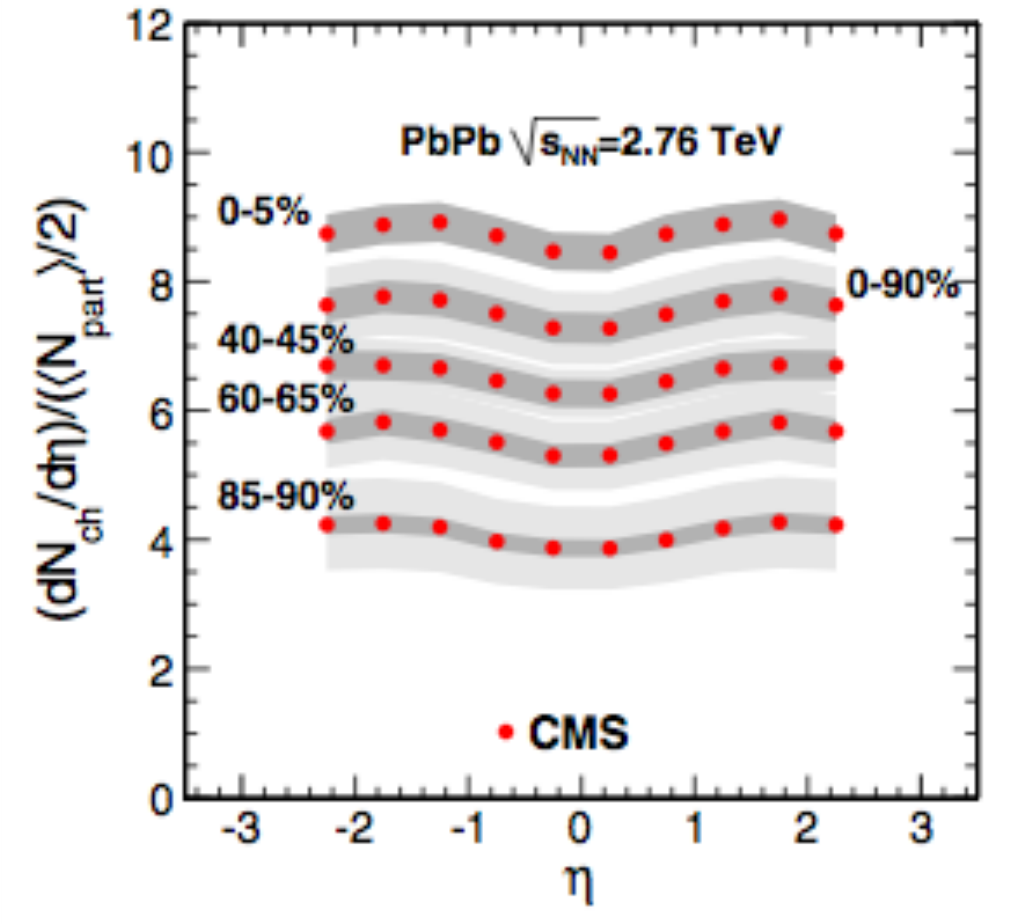}
\caption{Left panel shows the pseudorapidity multiplicity distribution at mid rapidity
 compared to various model predictions~\cite{ALICE_mult}. The right
 panel shows charged particle multiplicity density per participant
 pair for different centralities as a function of $\eta$~\cite{CMS_mult}.
}
\label{fig:dndeta}
\end{center}
\end{figure} 

\begin{figure}[ht]
\begin{center}
\includegraphics[width=0.9\columnwidth]{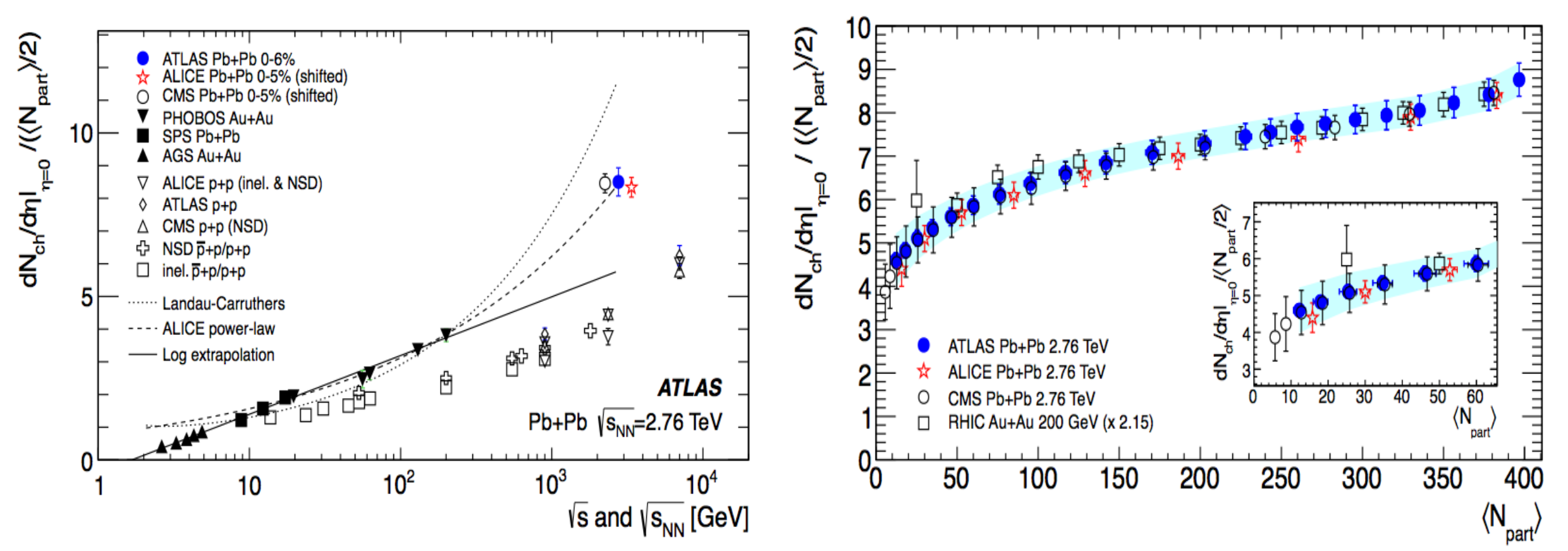}
\caption{Left panel shows the energy dependence of charged particle pseudorapidity density,
dN$_{\rm ch}$/d$\eta$ 
per colliding nucleon pair (0.5N$_{\rm part}$)
for pp (or p$\bar{\rm p}$ and AA collisions. The right panel shows
the centrality dependence of dN$_{\rm ch}$/d$\eta$  per participant
pair~\cite{ATLAS_mult}.
}
\label{fig:multiplicity}
\end{center}
\end{figure}

Results on charged particle multiplicity measurements were much awaited
and were the first ones to be available from the LHC
heavy ion run~\cite{ALICE_mult}. The value of charged particle 
multiplicity density (d$N_{\rm ch}/$d$\eta$) at central rapidity 
for \mbox{Pb-Pb} collisions at $\sqrt{s_{NN}}=
2.76$~TeV has been measured to be $\sim$1600 for central
collisions~(0-5\%). This is shown in the left panel 
of Fig.~\ref{fig:dndeta}. Also shown are recent predictions from
various models. As we see, the experimental result is on the higher
side of most of the predicted values. The rapidity distributions of
the charged particle multiplicity density per participant pair 
have been measured by the CMS experiment~\cite{CMS_mult} and 
shown in the right panel of Fig.~\ref{fig:dndeta} for peripheral
(85-90\%) to central (0-5\%) collisions. 

Energy dependence of charge particle multiplicity density has been
compiled by combining the results from all three LHC experiments along
with those from RHIC, SPS and AGS. This is shown in the left panel of 
Fig.~\ref{fig:multiplicity} for central collisions. 
The results for corresponding \mbox{p-p} 
collisions for NSD processes are also superimposed in the figure. 
We note here two interesting facts: (1) the energy dependence is
steeper for heavy ion collisions than for \mbox{p-p} collisions, and
(2) a significant increase in the pseudorapidity density at LHC compared to
extrapolations from lower energies.

The right panel of Fig.~\ref{fig:multiplicity} shows
centrality dependence of 
charged particle multiplicity density per participant pair 
obtained by all three LHC experiments along with the scaled value for \mbox{Au-Au}
collisions at RHIC top energy. A moderate variation~\cite{ATLAS_mult}
is observed, from a value of 4.6 for peripheral collisions to 8.8 for
central collisions. The RHIC data, scaled by a factor of 2.15 are seen
to match well with those of the LHC results.

\subsection{Size of the fireball and lifetime}

\begin{figure}[t]
\begin{center}
\includegraphics[width=0.95\columnwidth]{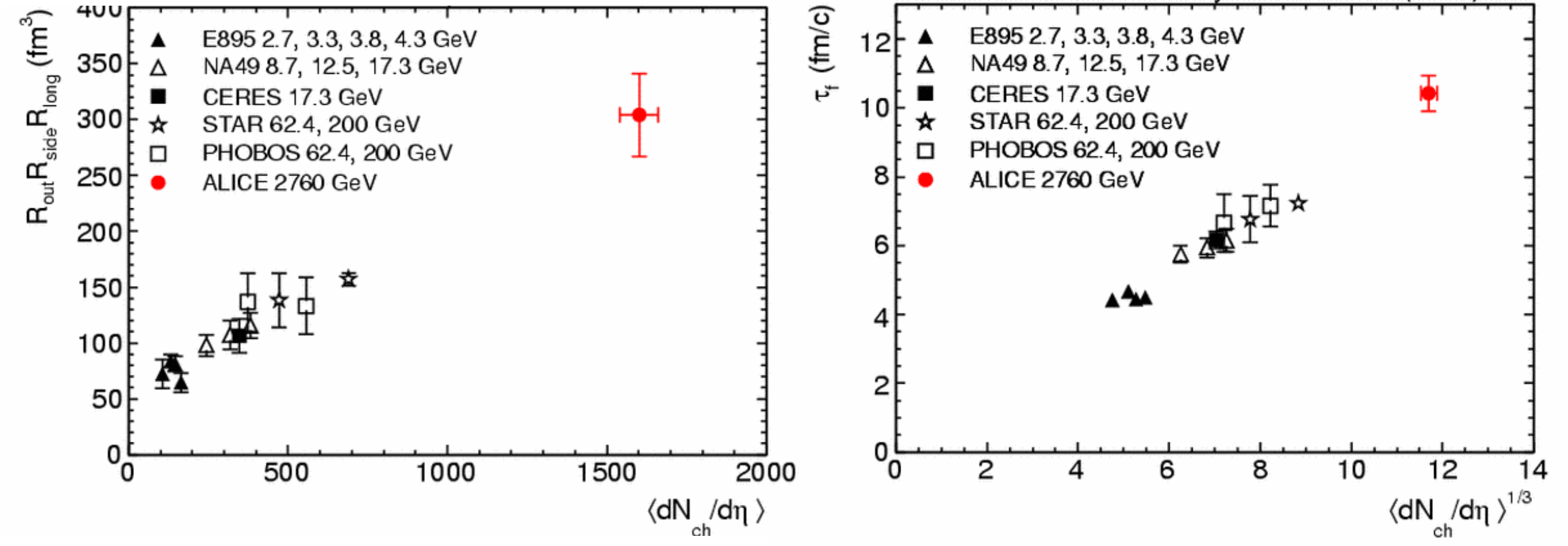}
\caption{ 
Left Panel shows the local freeze-out volume measured by identical 
pion interferometry as a function of pseudo-rapidity density of 
charged particles from AGS, SPS, RHIC and LHC energies~\cite{ALICE_hbt}. The right 
panel shows the system lifetime ($\tau_f$) for these energies~\cite{ALICE_hbt}
}
\label{fig:hbt}
\end{center}
\end{figure}

The information about the freeze-out volume and lifetime of
the created system in \mbox{p-p} and 
heavy ion collisions from the measured particle momenta have been
extracted~\cite{ALICE_hbt} by using 
the method of Bose-Einstein correlations or the
Hanbury-Brown and Twiss (HBT) technique. 
All the three pion source radii, longitudinal ($R_{\rm long}$), 
outward ($R_{\rm out}$), and sideward ($R_{\rm side}$) are observed to
grow compared to those of the results from RHIC.
Fig~\ref{fig:hbt} shows the product of the three radii 
as a function of charged particle
multiplicity density for central collisions at LHC, RHIC and SPS
energies. The homogeneity volume is observed to be larger by a factor of
two at LHC compared to RHIC~\cite{debasish}.

or the decoupling time ($\tau_f$), which is the
time between collision and freeze-out for hadrons at mid-rapidity, has been
estimated~\cite{ALICE_hbt} from $R_{\rm long}(k_{\rm T})$ by using the
expressions:
\begin{equation}
R_{\rm long}^2(k_{\rm T}) = \frac{\tau_f^2 T}{m_{\rm T}}
\frac{K_2(m_{\rm T}/T)}{K_1(m_{\rm T}/T)}, 
\end{equation}
where $m_{\rm T}$ is the pion mass, $T$ is the kinetic freeze-out
temperature ($\sim$0.12 GeV) and $K_1$ and $K_2$ are the integer order
modified Bessel functions. The decoupling times, shown in
Fig.~\ref{fig:hbt}, are seen to be 
The decoupling time for mid-rapidity pions, as shown in
Fig.~\ref{fig:hbt} are seen to exceed 10 fm/$c$ which is 40\% larger than at RHIC.

\subsection{Anisotropic flow}

Anisotropic flow measurement in nuclear collisions provides one of the most important measures of the
collective dynamics of the system ~\cite{ALICE_flow,raimond,jurgen}.
The magnitude of the anisotropic flow depends strongly on the
friction in the created matter, 
characterized by the shear viscosity over entropy density ratio ($\eta/s$).
The large elliptic flow observed at RHIC provided compelling evidence
for strongly interacting matter which, in addition, 
appears to behave like an almost perfect fluid~\cite{kovtun}.

\begin{figure}[ht]
\begin{center}
\includegraphics[width=0.45\columnwidth,height=4.3cm]{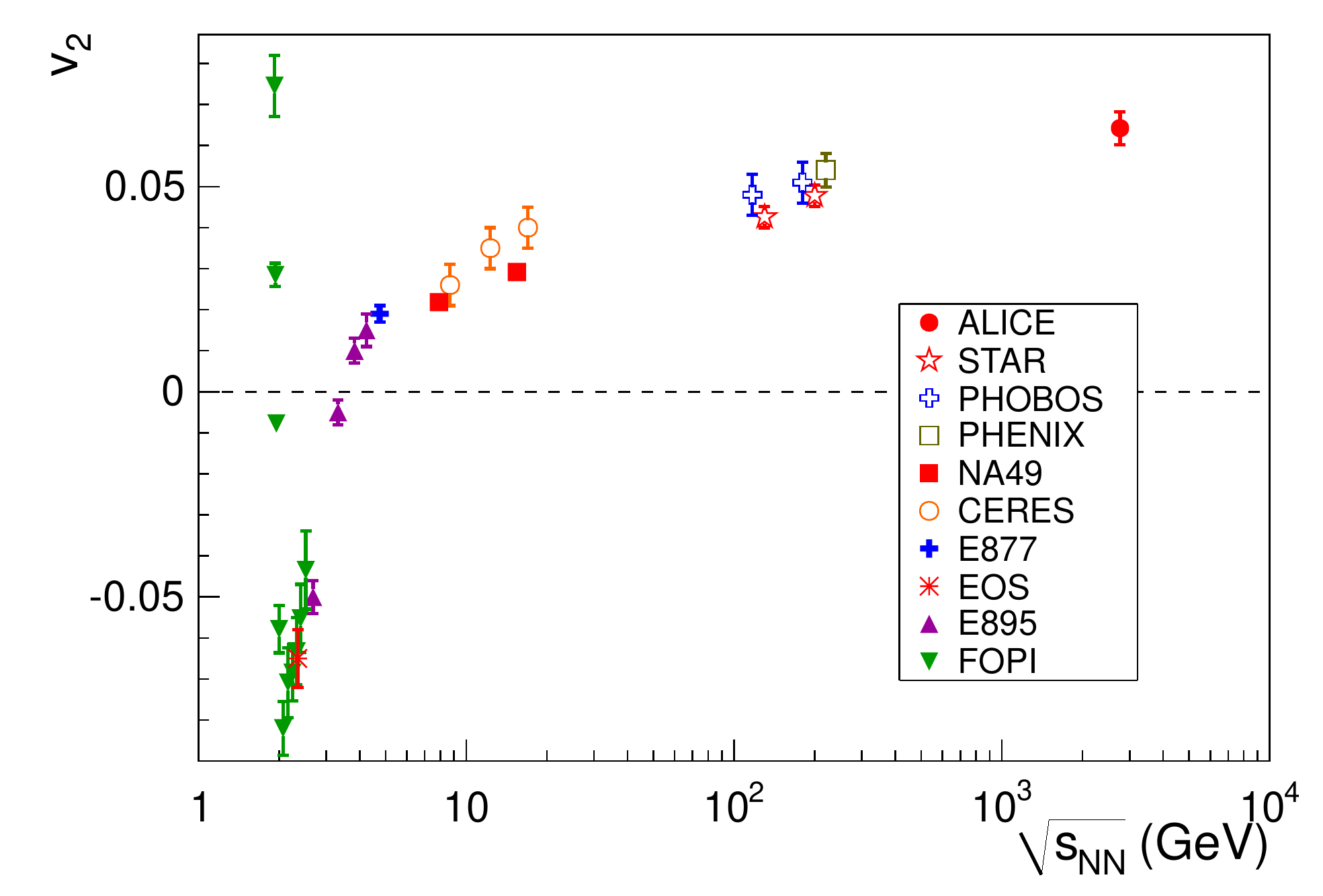}
\includegraphics[width=0.45\columnwidth,height=4.3cm]{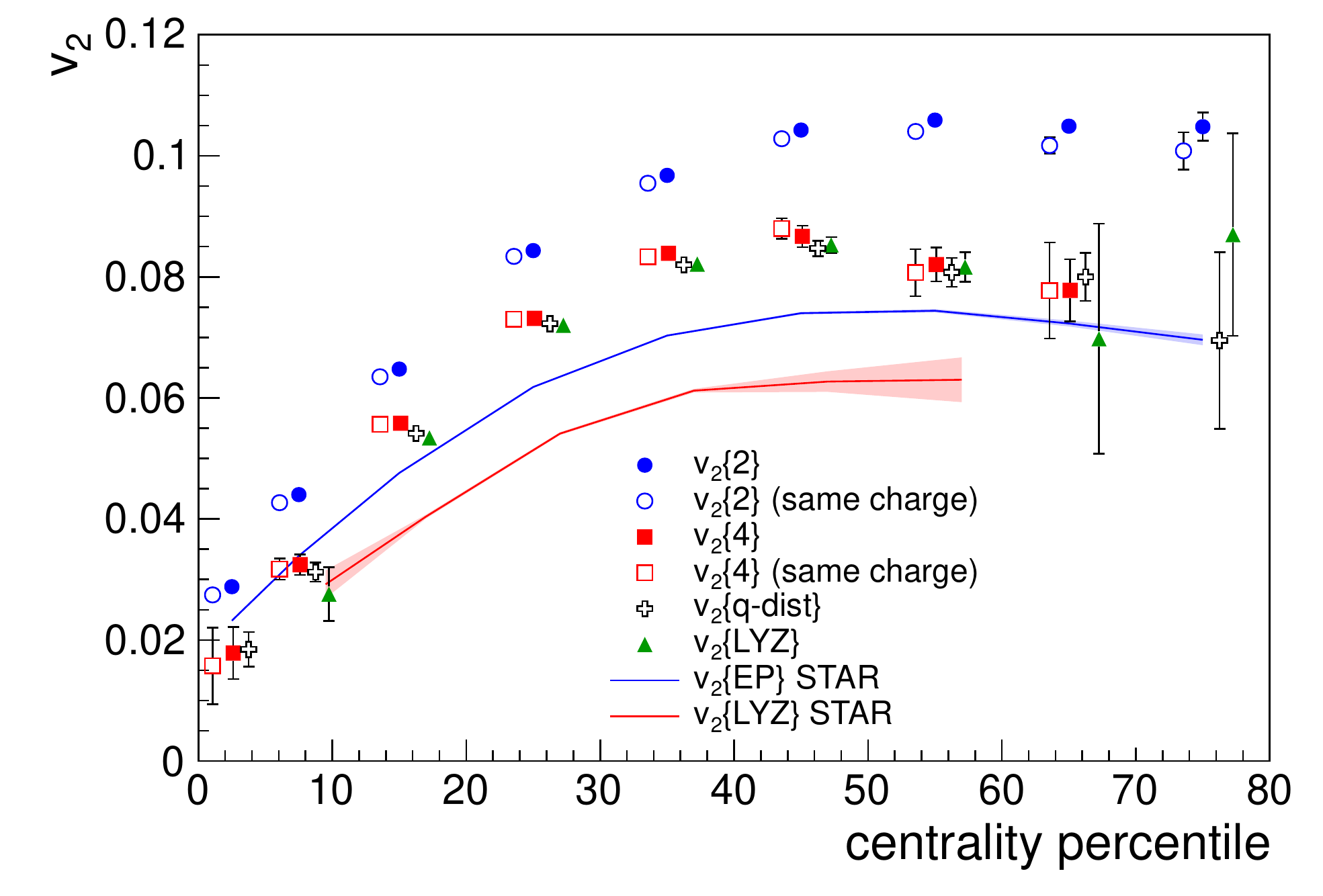}
\caption{
Integrated elliptic flow ($v_2$) a function of collision energy for
non-central collisions (left) and as a function of centrality
going from central to peripheral collisions (right)~\cite{ALICE_flow}.
}
\label{fig:flow}
\end{center}
\end{figure} 

The left panel of FIg.~\ref{fig:flow} shows the integrated elliptic 
flow measured in the 20-30\%
centrality class, plotted for LHC energy along with results from RHIC
and lower energies. An increase
in the magnitude ($\sim$30\%) 
of $v_2$ at LHC has been observed compared to the results at RHIC top
energy. This increase may be the result of the 
increase in the average transverse momentum at LHC
compared to RHIC.
The right panel of Fig.~\ref{fig:flow} 
shows the centrality dependence of elliptic flow ($v_2$) for \mbox{Pb-Pb} collisions
at $\sqrt{s_{NN}}=2.76$~TeV.
The integrated elliptic flow increases from central to peripheral
collisions and reaches a maximum value in the 50-60\% and 40-50\% 
centrality classes. These results are consistent with hydrodynamic
model calculations. 

\subsection{Fluctuations and correlations}

Event-by-event studies of fluctuations and correlations 
provide direct evidence for the QGP formation and of the thermalized
system formed in heavy ion collisions. The order of phase transition
may also be inferred from these studies. At the LHC, the first such studies have been
performed for net-charge and mean-\pt fluctuations. 

\begin{figure}[tbh]
\begin{center}
\includegraphics[width=0.47\columnwidth,height=4.3cm]{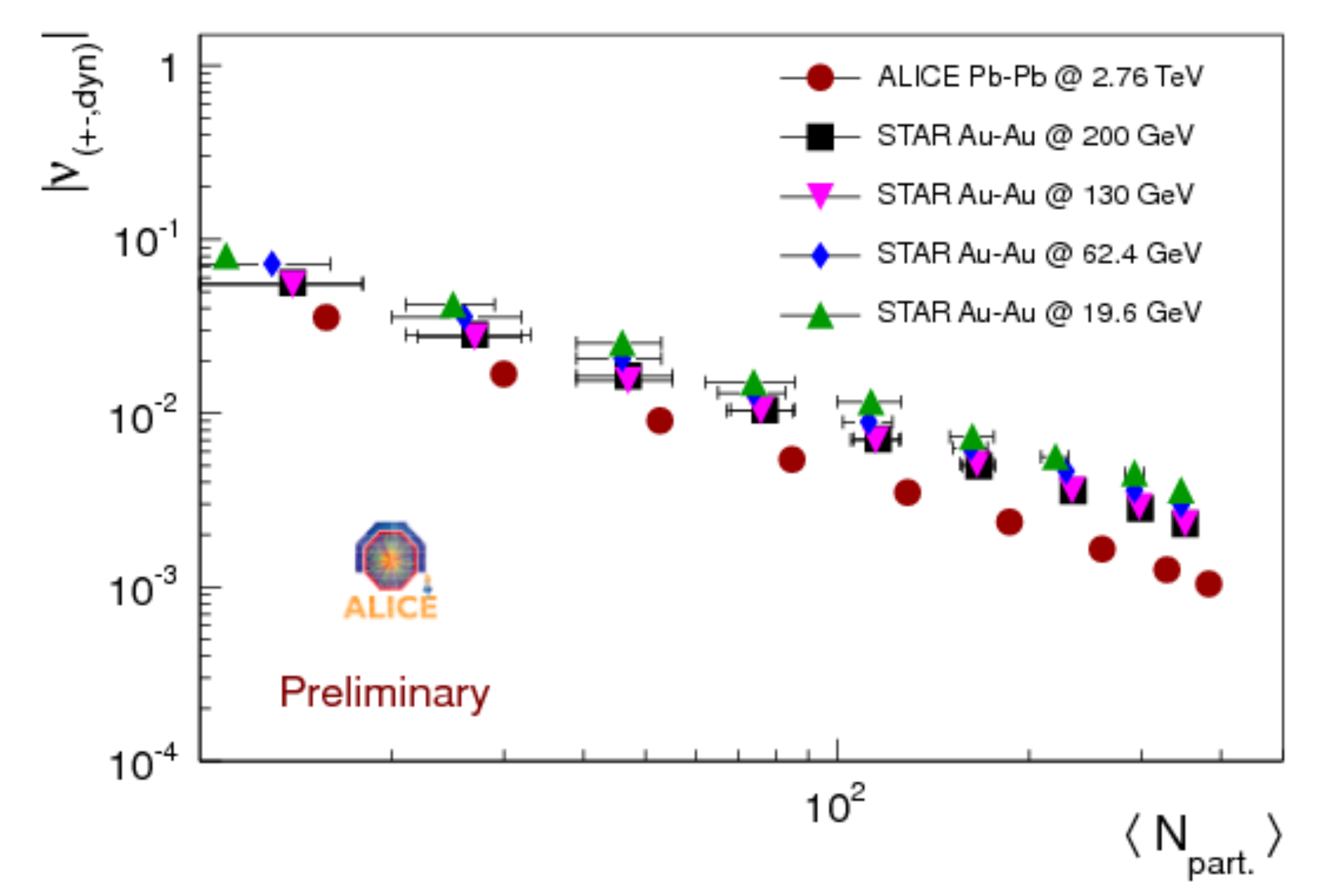}
\includegraphics[width=0.47\columnwidth,,height=4.3cm]{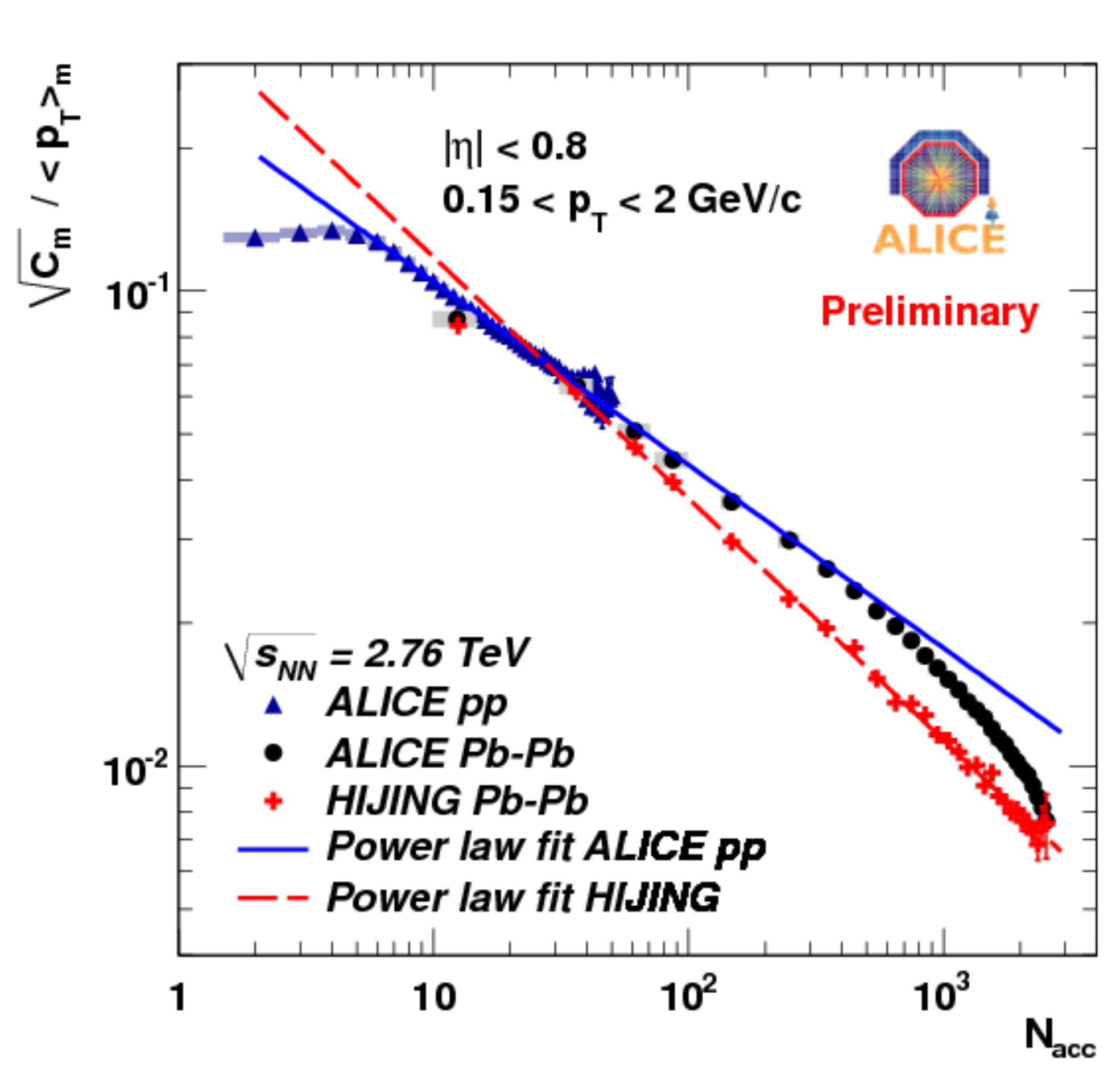}
\caption{
(Left) Dynamical net--charge fluctuations, \nudyn, of charged particles  
as a function of centrality, expressed by the  
number of participating nucleons, for  
Pb--Pb (right) events at $\sqrt{s_{NN}} = 2.76$~TeV, along with data 
from RHIC~\cite{StarChargeFluctuations}.
(Right) The dependence of the relative mean \pt fluctuations on the number of  
accepted tracks in \mbox{Pb-Pb} collisions, compared to different 
model predictions~\cite{panos,satya}.
}
\label{fig:fluc}
\end{center}
\end{figure}

The fluctuations of net-charge (difference of +ve to -ve charge
multiplicities) depend on the squares of the charge states present in the system. 
Fluctuations in a QGP phase having quarks (with fractional charges) as the charge carriers are
significantly different from those of a hadron gas (with unit charges). 
Net-charge fluctuations may be expressed by the quantity, 
$D = 4 \frac{\langle \delta Q^2\rangle}{N_{\rm ch}}$,
where $\langle \delta Q^2 \rangle$ is the variance of the 
net--charge $Q$ with $Q = N_+ - N_-$ and $N_{\rm ch} = N_+ + N_-$. Here 
$N_+$ and $N_-$ are the numbers of positive and negative particles.
Typical values of $D$ are close to 1 for a QGP and 3 for a hadron
gas ~\cite{JeonKoch00}. Thus a measured value of $D$ will be able to
indicate whether the particles originate from a QGP or from hadron gas.
Experimentally the dynamic fluctuations in net-charge may be expressed in terms of
\nudyn~\cite{pruneau}, which is related to $D$. The left panel of
Fig.~\ref{fig:fluc}  gives the centrality dependence of the absolute
value of \nudyn in a log-log scale. The measurements at LHC
are compared to the corresponding
measurements ~\cite{StarChargeFluctuations} at RHIC, which show
an additional reduction of the magnitude of fluctuations at LHC.

The mean transverse momentum $<$$p_{\rm t}$$>$  
of emitted particles in an event is
correlated to the temperature associated with the \pt distribution and thus
to the transverse collective expansion of the colliding system. 
The study of the $<$$p_{\rm t}$$>$  
fluctuations can probe the dynamics and
the underlying correlations of the created system.  
A  two particle correlator,  $C_m = \langle \Delta p_{\rm{t},i} , 
\Delta p_{\rm{t},j} \rangle_m$ is used which comprises of the
dynamical component of the relevant fluctuations. 
The relative fluctuation expressed as
$\sqrt{C_{m}}/ \langle p_{\rm{t}} \rangle_{m}$, 
is shown in the right panel of Fig.~\ref{fig:fluc} for 
\mbox{Pb-Pb} collisions at $\sqrt{s_{NN}} = 2.76$~TeV as a function of
number of accepted tracks $(N_{\rm acc}$). Compared to the \mbox{p-p}
reference, the data points for central collisions 
(higher $N_{\rm acc}$)  show reduction of fluctuations. 

\section{Quarkonia and heavy flavour} 

Comparison of the suppression pattern of quarkonia (flavourless mesons whose constituents are a quark
and its own anti-quark) in nuclear collisions to those of the \mbox{p-p}
collisions and the study of heavy-flavour particles which are
abundantly produced at LHC energies, 
provide detailed information about the early conditions of the produced fireball.

\begin{figure}[ht]
\begin{center}
\includegraphics[width=0.47\columnwidth,,height=4.7cm]{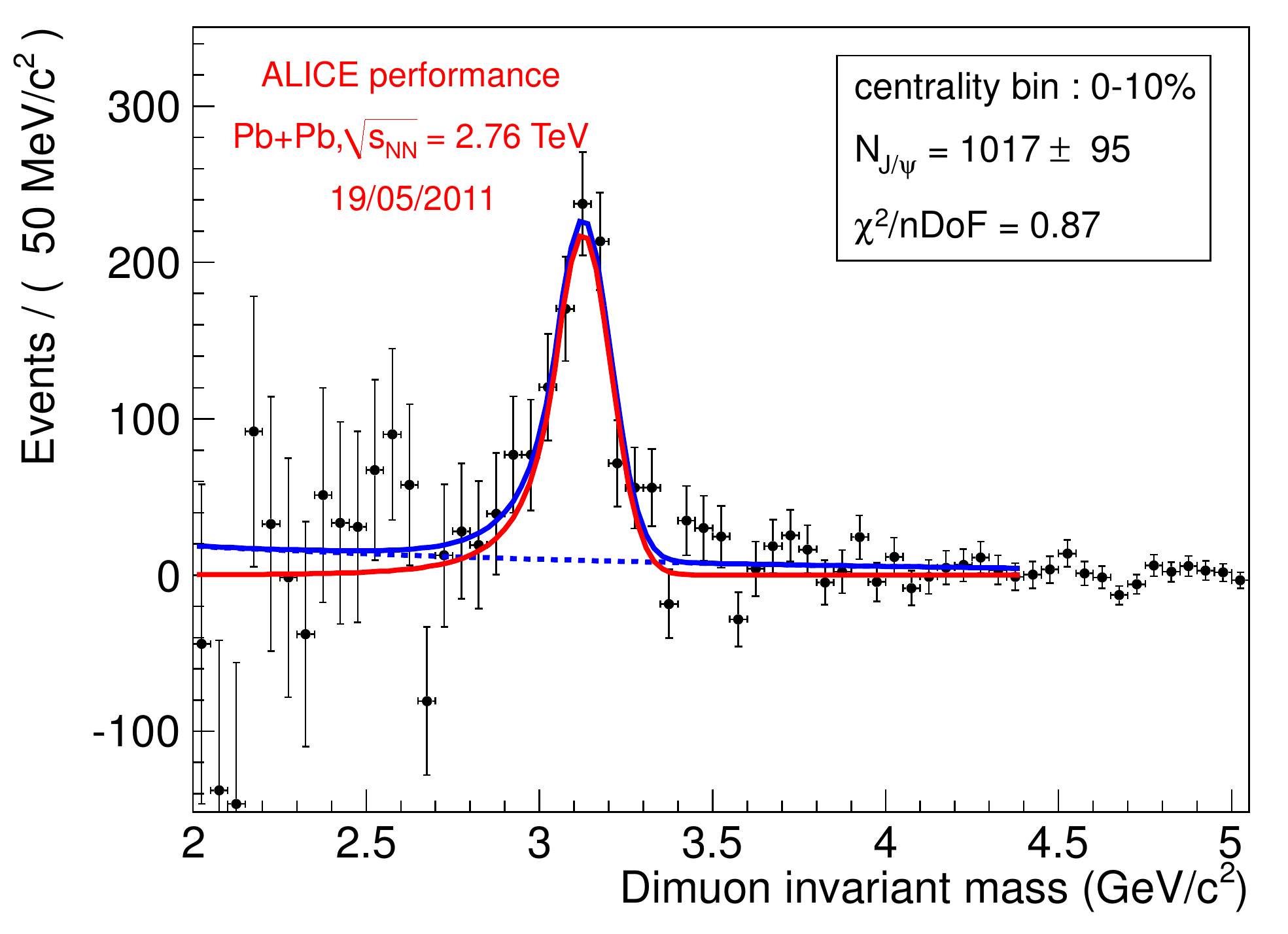}
\includegraphics[width=0.47\columnwidth,height=4.9cm]{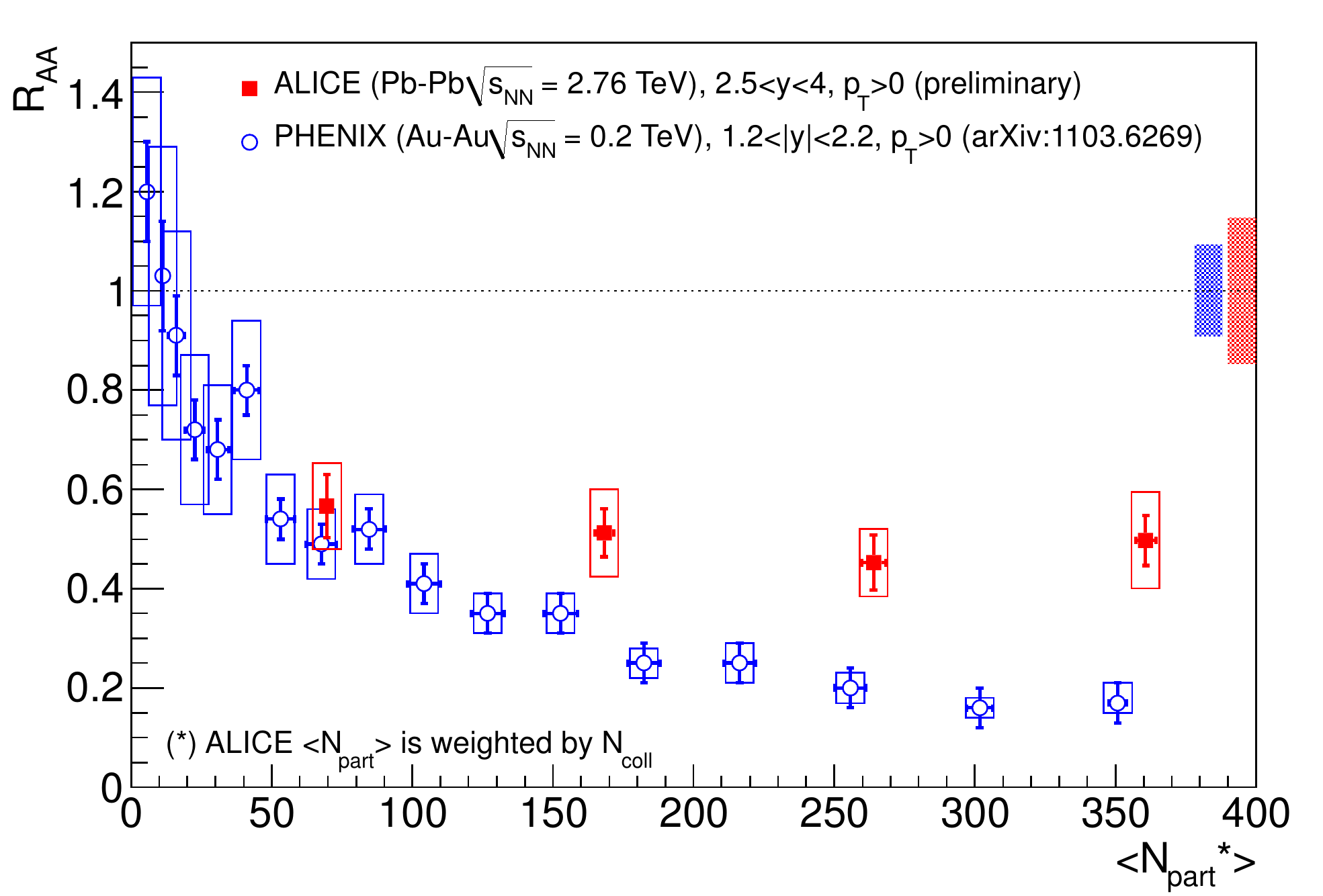}
\caption{Left panel shows the background subtracted 
invariant mass distribution for opposite-sign muon pairs in
  the centrality class 0-10\%, showing the  $J/\psi$ peak. The right
  panel shows
$R_{\rm AA}$ of $J/\psi$ as a function of number of participants for
LHC data compared to those of the data from RHIC. Figures are taken 
from~\cite{ALICE_quarkonia1,ALICE_quarkonia2}.
}
\label{fig:quarkonia}
\end{center}
\end{figure}

\begin{figure}[ht]
\begin{center}
\includegraphics[width=0.47\columnwidth,height=4.7cm]{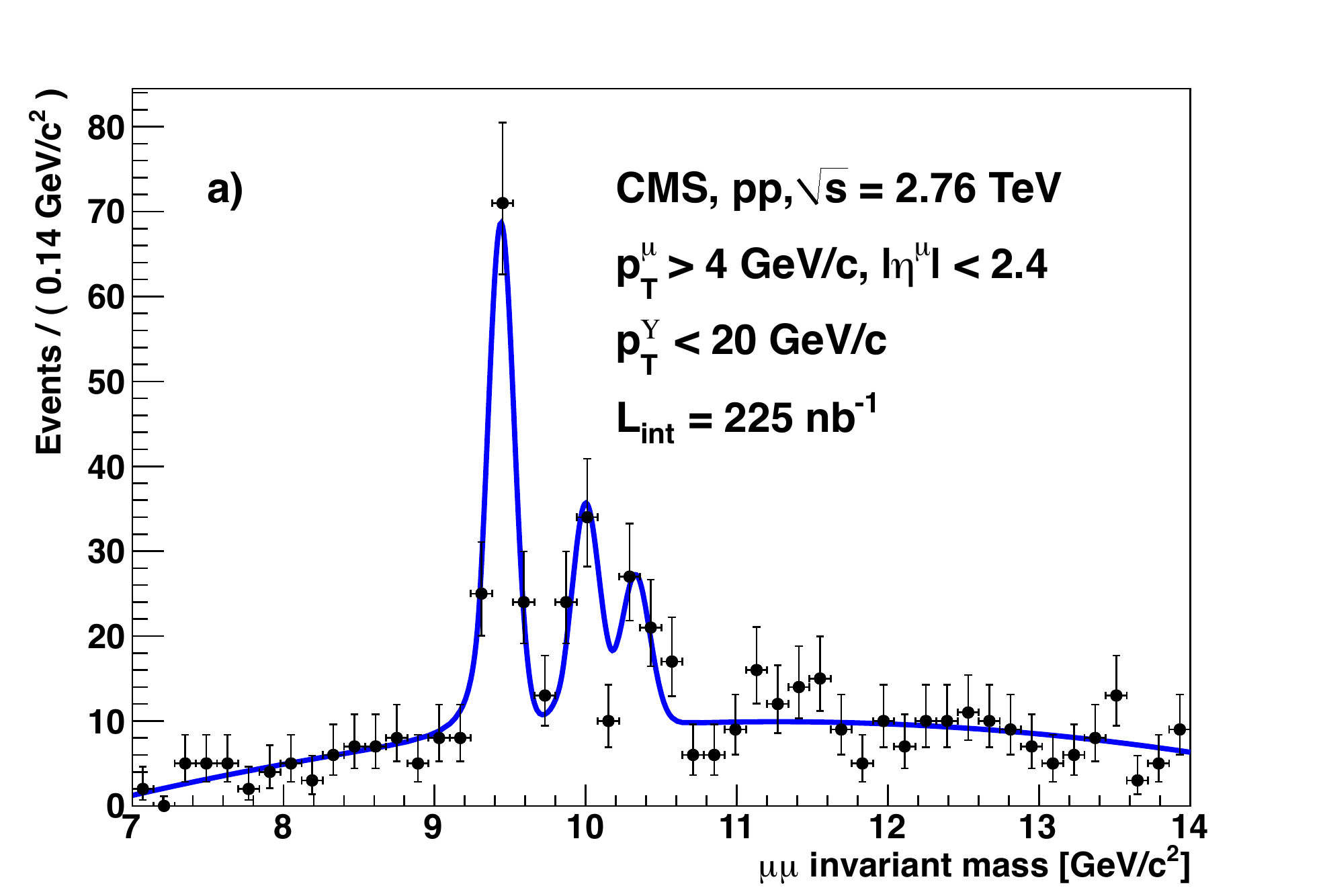}
\includegraphics[width=0.47\columnwidth,,height=4.7cm]{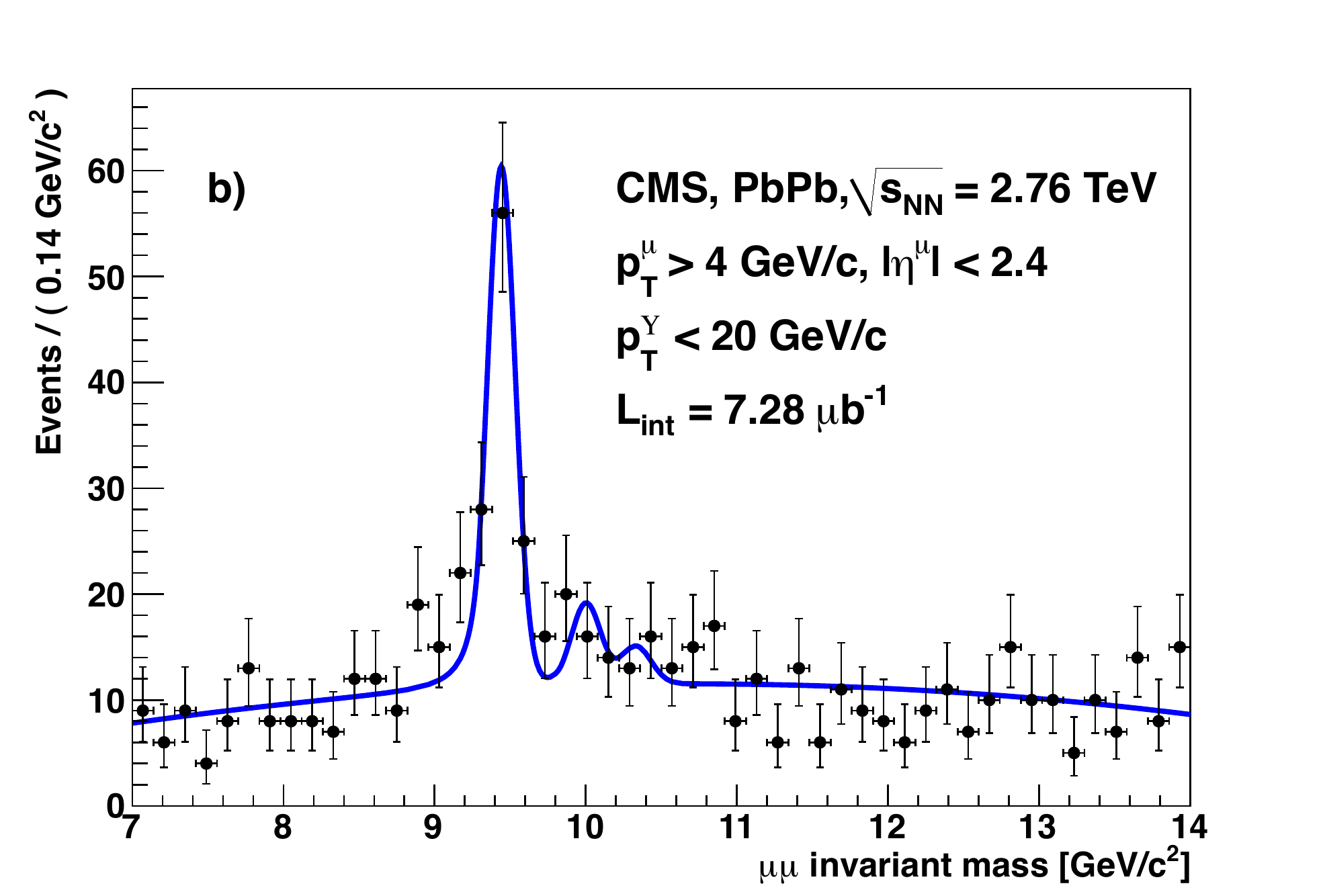}
\caption{Dimuon invariant-mass distributions in the $\Upsilon$
 region~\cite{CMS_quarkonia} for \mbox{p-p} (left panel) and 
 \mbox{Pb-Pb} (right panel) collisions. The solid lines are fits to extract the yields.
}
\label{fig:quarkonia2}
\end{center}
\end{figure}

Opposite sign dimuon invariant mass distribution, after the 
subtraction of the combinatorial background using the event-mixing 
technique shows the $J/\psi$ peak as depicted in  
the left panel of Fig.~\ref{fig:quarkonia}. 
The nuclear modification of factor $R_{\rm AA}$ which gives the
deviation in $J/\psi$ yields from \mbox{A-A} collisions related to the scaled
yields of $J/\psi$ from \mbox{p-p} collisions, is shown in the right
panel of figure. The $R_{\rm AA}$ is seen to be
about 0.5 and independent of centrality. 
The suppression of $J/\psi$ is seen to be about 
factor 2 less at LHC compared to the results at RHIC from
PHENIX experiment.  On the other hand,
the ATLAS and CMS results~\cite{CMS_quarkonia} show that
$J/\psi$ is strongly suppressed at high \pt, and similar to the PHENIX
data at RHIC. In order to understand the suppression and regeneration
effects, a better knowledge of the cold nuclear matter effects is
required. For this reason a \mbox{p–Pb} run, which will address these
nuclear effects, is being planned in the year 2012.

First measurements of $\Upsilon$ production with \mbox{p-p}
and \mbox{Pb–Pb} collisions at LHC~\cite{CMS_quarkonia} are made and
shown in Fig.~\ref{fig:quarkonia2}. The higher state contribution
relative to the ground state is strikingly smaller in \mbox{Pb-Pb} collisions
compared to the \mbox{p-p} collisions. Even with the large statistical errors,
we can conclude that the excited, higher mass $\Upsilon$ states are more suppressed
than the tightly bound ground state $\Upsilon$(1S), 
as expected in a deconfinement scenario. 

\begin{figure}[ht]
\begin{center}
\includegraphics[width=0.47\columnwidth,height=4.7cm]{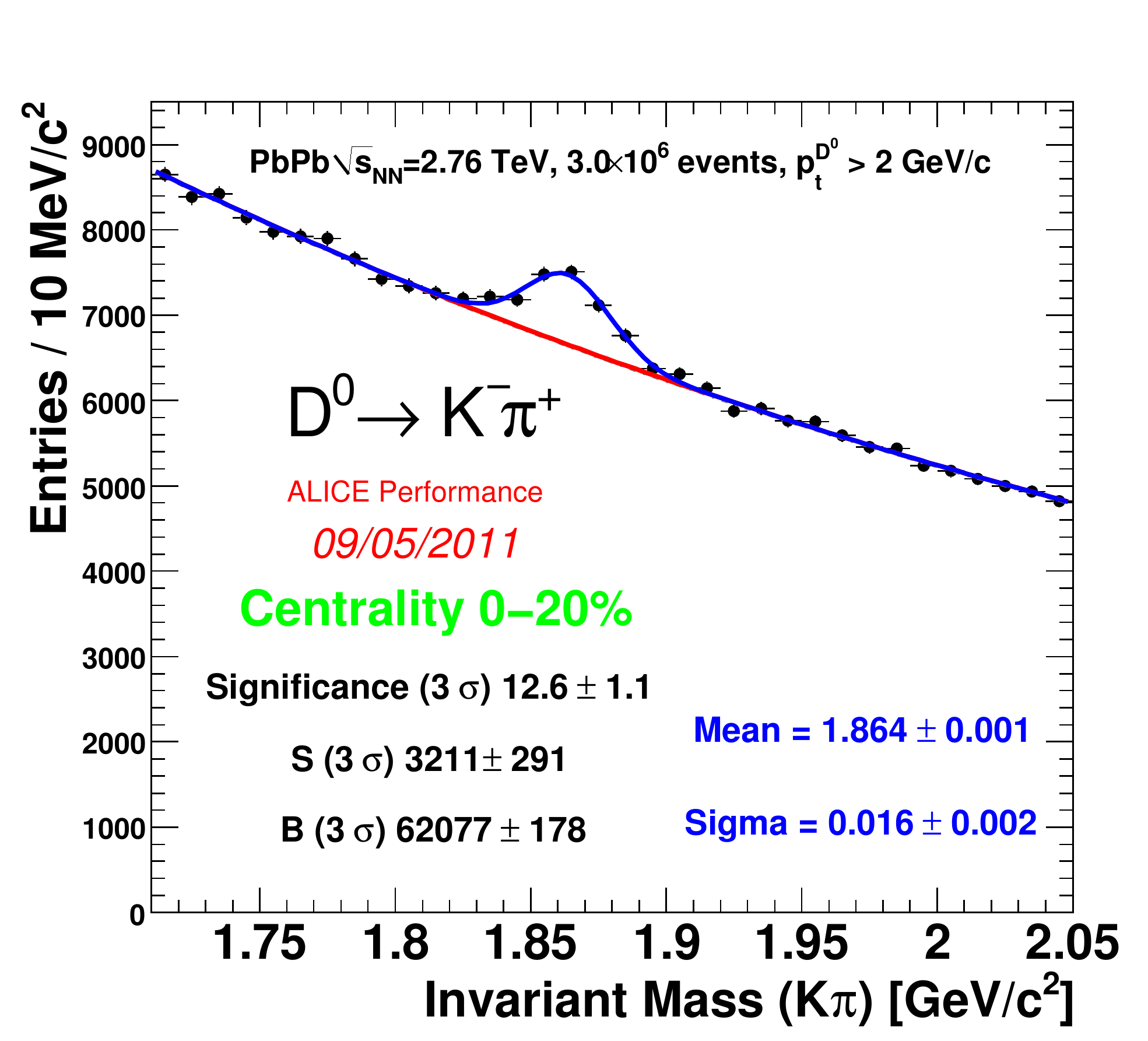}
\includegraphics[width=0.47\columnwidth,,height=4.7cm]{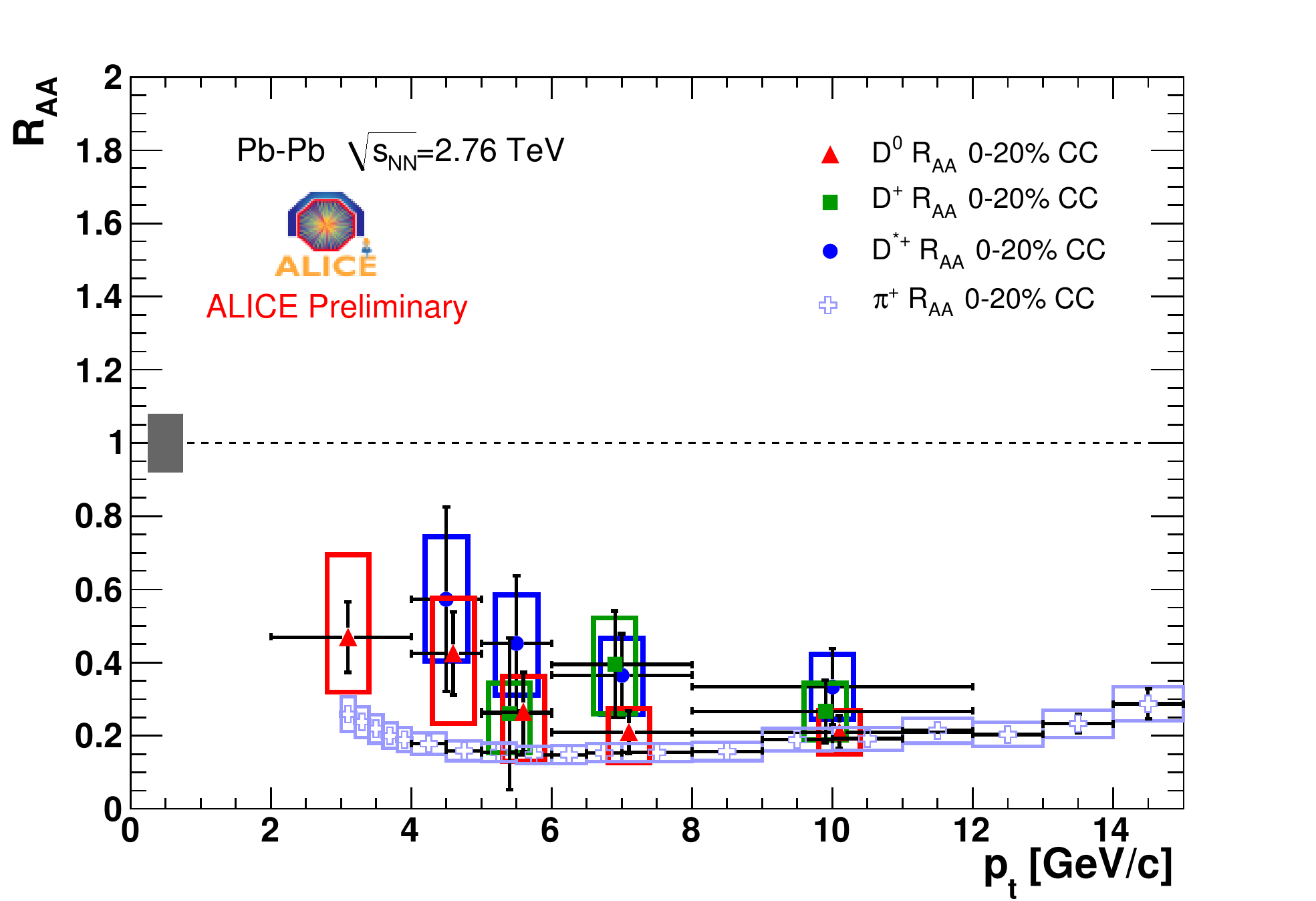}
\caption{(Left) $K\pi$ invariant mass showing the 
D$^0 \rightarrow K^- \pi^+ $ signal for central (0–20\%) 
\mbox{Pb–Pb} collision, and (Right) 
$R_{\rm AA}$ for D$_0$, D$_+$ and $\pi^+$ 
in central collisions~\cite{ALICE_dainese}.
}
\label{fig:heavyflavor}
\end{center}
\end{figure}

Heavy-flavours  such as D mesons are sensitive to energy density.
Their 
detection strategy at central rapidity is based on
the selection of displaced-vertex topologies, i.e. separation of
tracks 
from the secondary vertex from those from the primary vertex, 
large decay length and good 
alignment between the reconstructed D meson momentum and flight-line. 
An invariant-mass analysis is then used to extract the raw signal
yield. Corrections are made for detector acceptance and PID,
selection and reconstruction efficiencies.

The D$^0 \rightarrow K^- \pi^+ $ signal 
was measured in five \pt bins in 2-12~GeV/$c$ and the
 D$^+ \rightarrow K^- \pi^+ \pi^+$ signal in three bins in 5-12~GeV/c. 
An example $K\pi$ invariant mass distribution for \pt $>$ 2 GeV/$c$ in the
0-20\% centrality class is shown in the left panel of Fig.~\ref{fig:heavyflavor}.
A clear peak at the $D^0$ mass is seen.
The nuclear modification factor ($R_{\rm AA}$) 
of prompt D$^0$ and D$^+$ mesons in central 
(0-20\%) Pb–Pb collisions is shown in the right panel of 
Fig.~\ref{fig:heavyflavor}.  A strong suppression is observed, 
reaching a factor 4 to 5 for \pt$>$5~GeV/c 
for both D meson species.
For comparison, the values for charged pions is superimposed in the
figure.
The suppression for D mesons is comparable with that of the charged pions.

\section{Electroweak Probes}

\begin{figure}[t]
\begin{center}
\includegraphics[width=0.45\columnwidth,height=4.7cm]{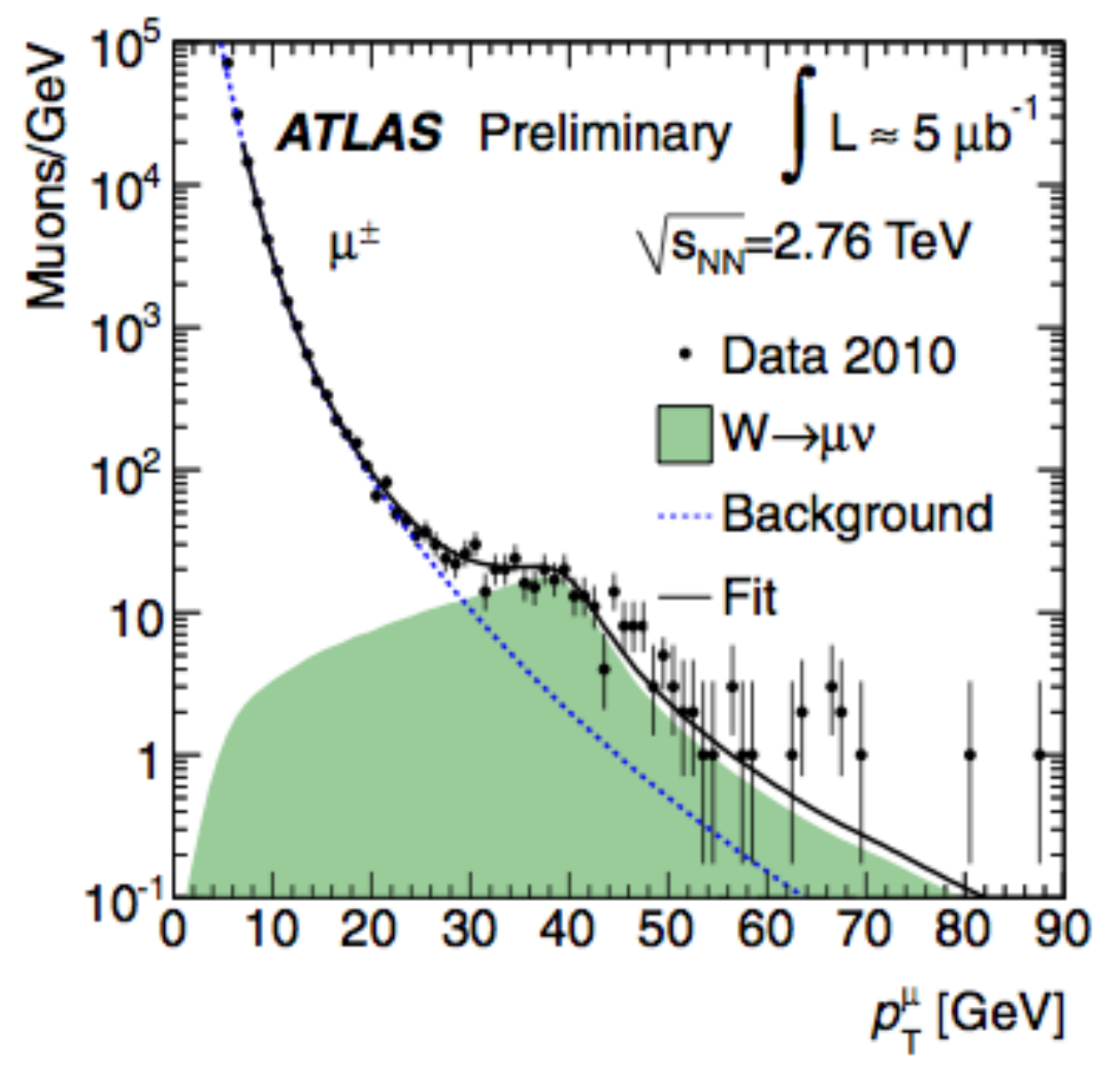}
\includegraphics[width=0.45\columnwidth,,height=4.7cm]{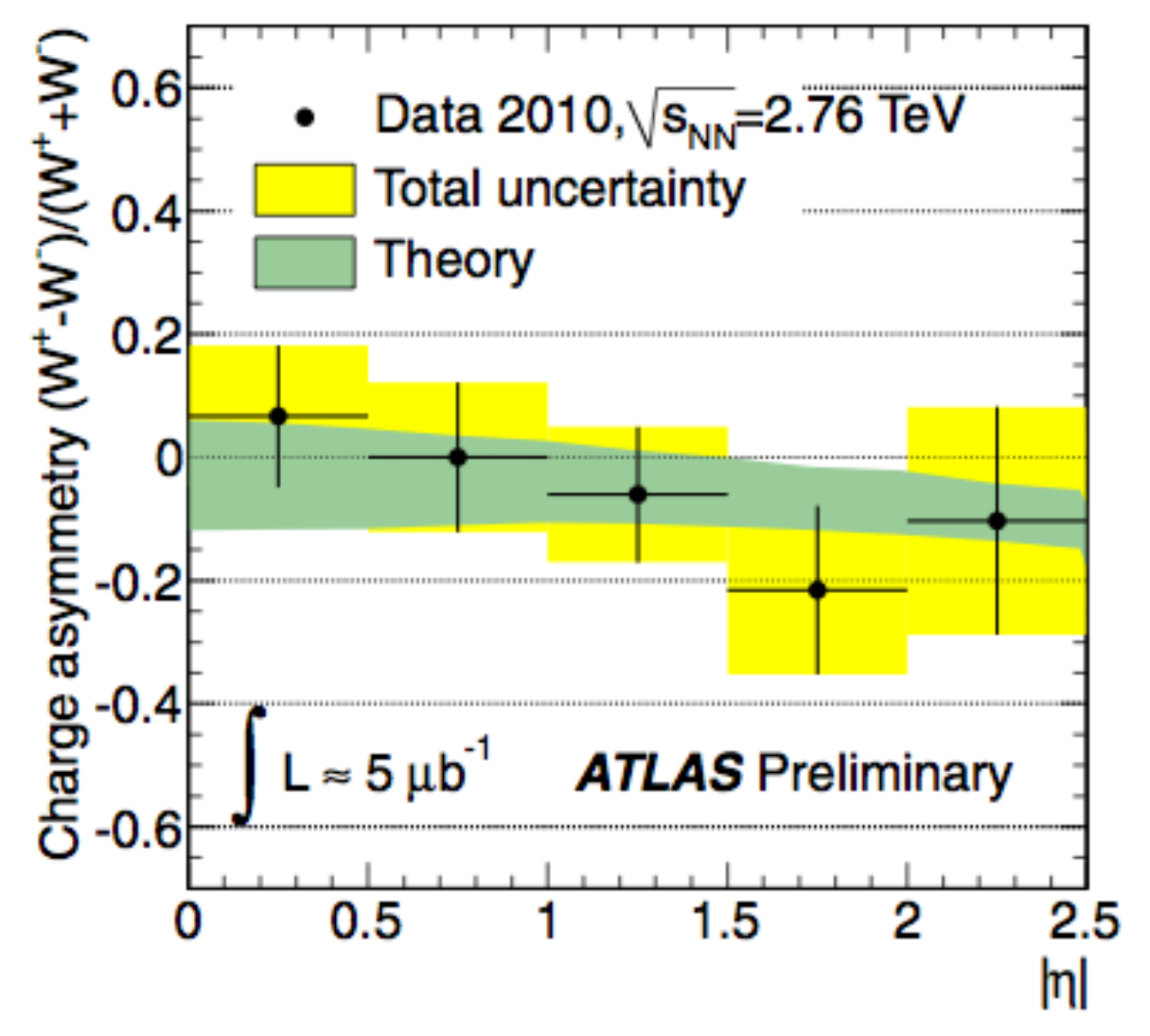}
\caption{(Left) Extraction of the number of 
$W\rightarrow \mu\nu$ events from the uncorrected inclusive  
muon \pt spectrum in   \mbox{Pb-Pb} collisions at $\sqrt{s_{NN}}= 
2.76$~TeV.
(Right) Muon charge asymmetry from $W^\pm$ decays. 
~\cite{ATLAS_weak}.
}
\label{fig:W}
\end{center}
\end{figure}

\begin{figure}[tbp]
\begin{center}
\includegraphics[width=0.45\columnwidth,height=4.5cm]{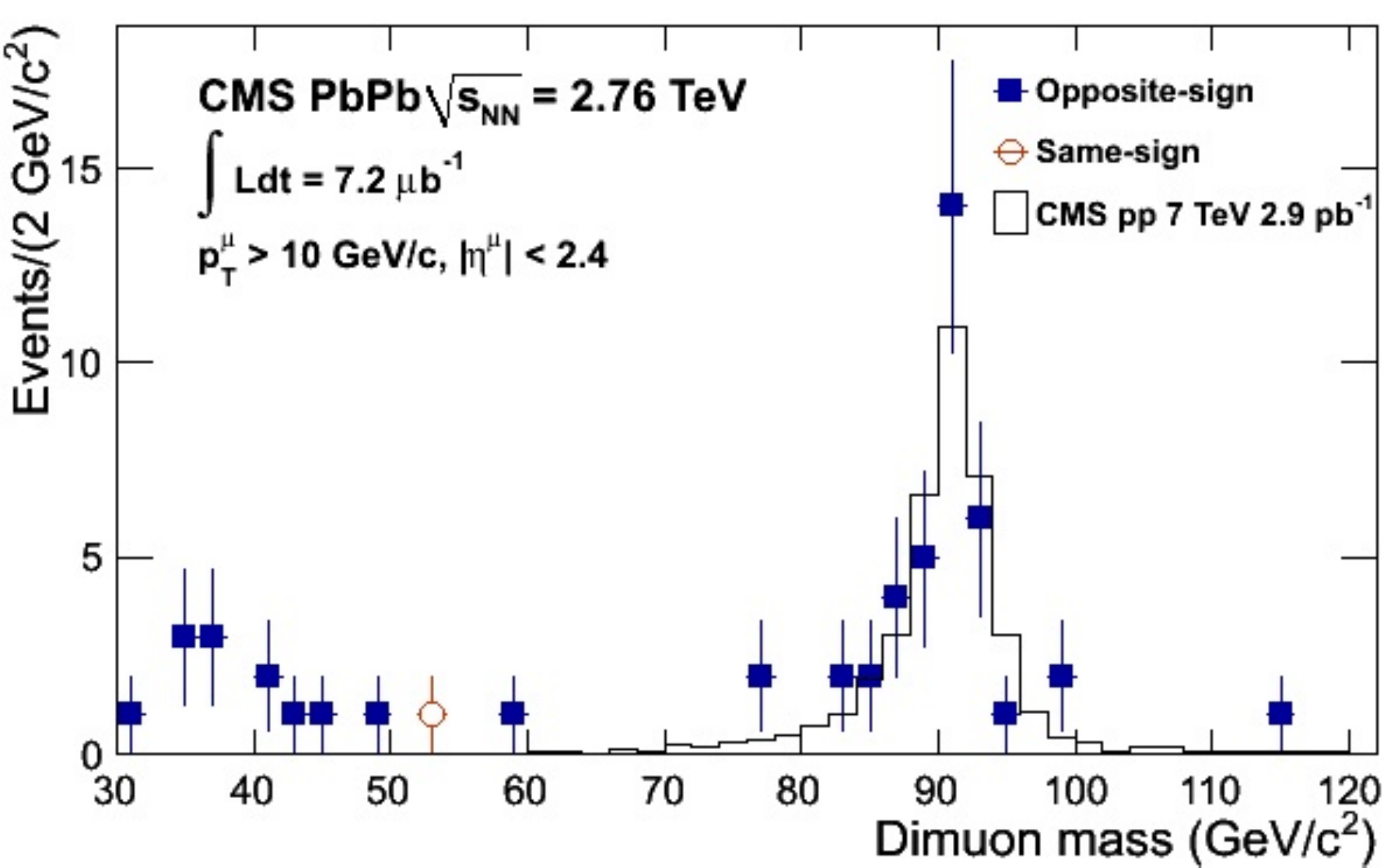}
\includegraphics[width=0.45\columnwidth,,height=4.7cm]{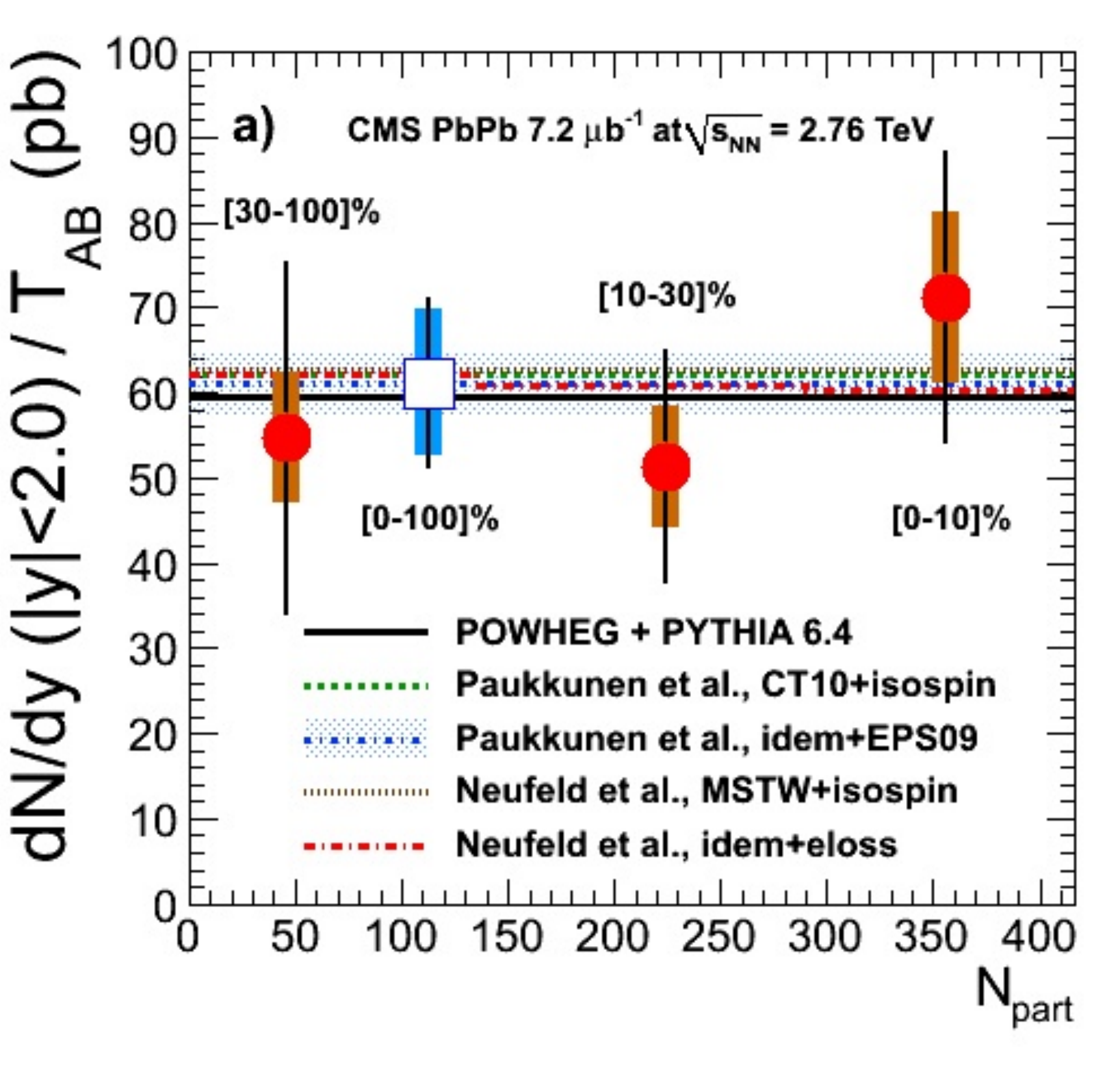}
\caption{(Left) Dimuon invariant mass showing a clear peak around
$Z$ boson mass in \mbox{Pb-Pb} collisions 
at $\sqrt{s_{NN}}= 2.76$~TeV. (Right) $Z$ differential yield divided by 
the nuclear overlap function as a function of number of participating nucleons.
~\cite{CMS_weak}.
}
\label{fig:Z}
\end{center}
\end{figure} 
 
The ATLAS and CMS experiments have measured the $W$ and $Z$ production
in the heavy ion collisions at the LHC~\cite{ATLAS_weak,CMS_weak}. 
ATLAS studies the production of $W$ bosons  via the measurement of the
inclusive muon \pt spectrum as shown in the left panel of Fig.~\ref{fig:W}. The spectrum
is fitted (solid line) with two components: signal $W\rightarrow\mu\nu$ (shaded
area) and a background parametrization (dashed line). The amplitude of
$W$-decay curve gives the number of $W$ events on the data. Similar
studies are also reported by the CMS experiment. A
precision measurement of the $W$ boson production charge asymmetry can
provide a test of PDF (parton distribution function), which may lead
to the understanding of possible modifications to the PDF due to
nuclear effects. The muon charge asymmetry is shown in the right
panel of Fig.~\ref{fig:W} for different $\eta$-window. Although the 
statistical errors are dominating, one may conclude that 
production cross section for $W^-$ and $W^+$ is similar.

Among the leptonic decays of the electroweak bosons, the study of the
$Z$ boson in the $\mu^+\mu^-$ channel is the cleanest
one~\cite{CMS_weak}. The reconstruction of the Z boson in the dimuon 
channel is done by requiring two opposite-charge muons, each with 
$p^{\mu}_{\rm t} > 10$~GeV/c and $|\eta_\mu|< 2.4$. A clear peak in
the 60-120 GeV/$c^2$ mass region is observed as shown in the invariant
mass spectrum in the left panel of Fig.~\ref{fig:Z}. Differential
yields divided by the nuclear overlap function is plotted in the right
panel of the figure, which shows that the distribution is flat (within
uncertainty) as a function of centrality.

\section{Hard probes}

Hard processes, where we study the higher side of the \pt
spectrum, provides a means to probe the conditions of the collision at very early times. 
This is typically done by means of suppression observables such as
R$_{\rm AA}$ which indicate deviations from the expected scaling of
yields by the number of binary collisions, estimated using the Glauber
model, and by the study of jets quenching.
Suppression of high \pt hadron yields result from energy loss
suffered by hard-scattered partons passing through the medium.
The parton energy loss provides fundamental
information on the thermodynamical and transport properties
of the traversed medium.
Below we discuss the current LHC
results on these aspects. 

\begin{figure}[hbt]
\begin{center}
\includegraphics[width=0.9\columnwidth]{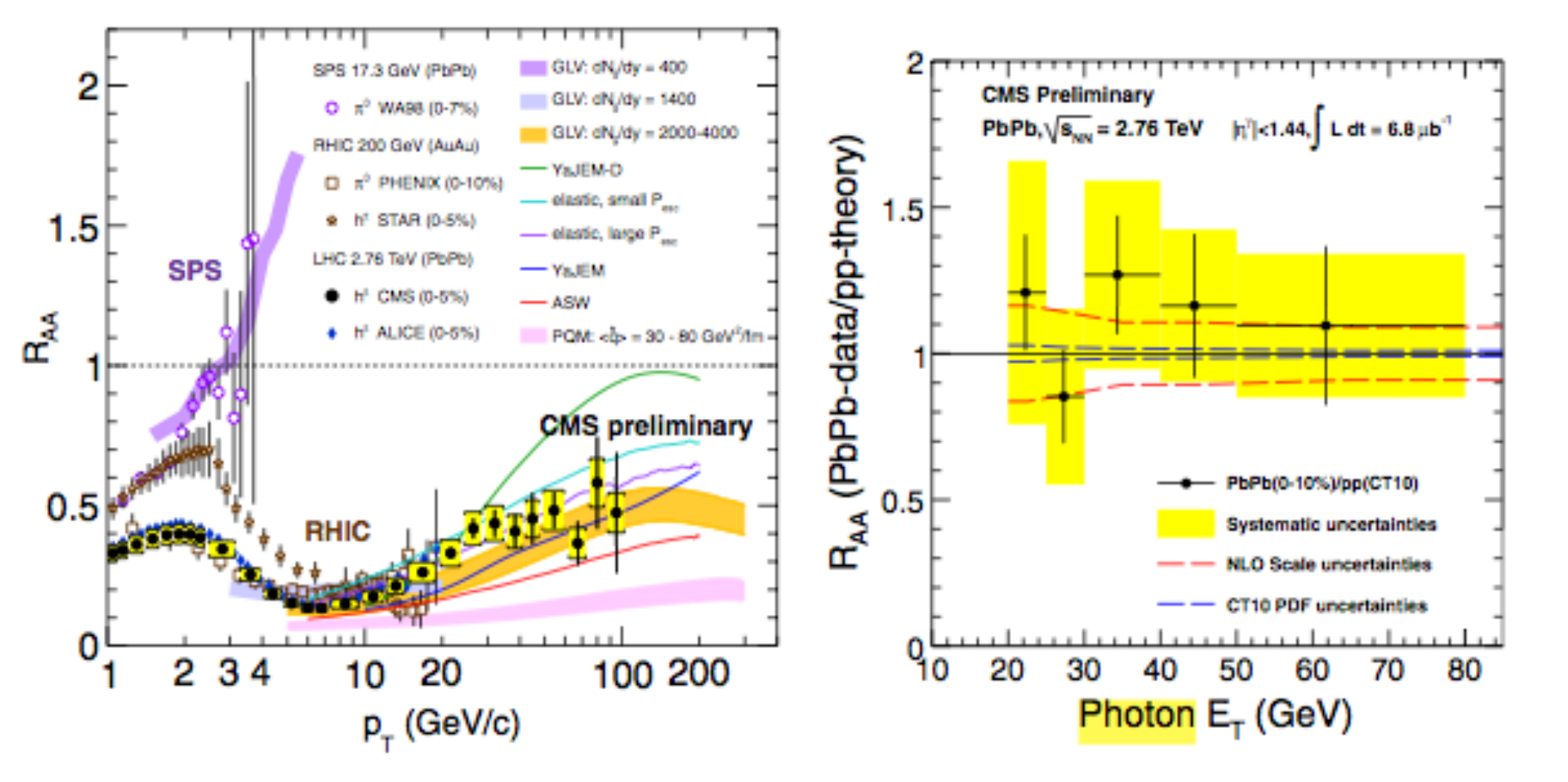}
\caption{(Left) $R_{\rm AA}$ as a function of \pt for neutral pions
and charged hadrons in central heay-ion collisions at SPS, RHIC and
LHC energies. (Right) $R_{\rm AA}$ of isolated photons as a function
of \pt for central events at LHC~\cite{CMS_mult}.
}
\label{fig:Raa}
\end{center}
\end{figure} 
\begin{figure}[h]
\begin{center}
\includegraphics[width=0.49\columnwidth]{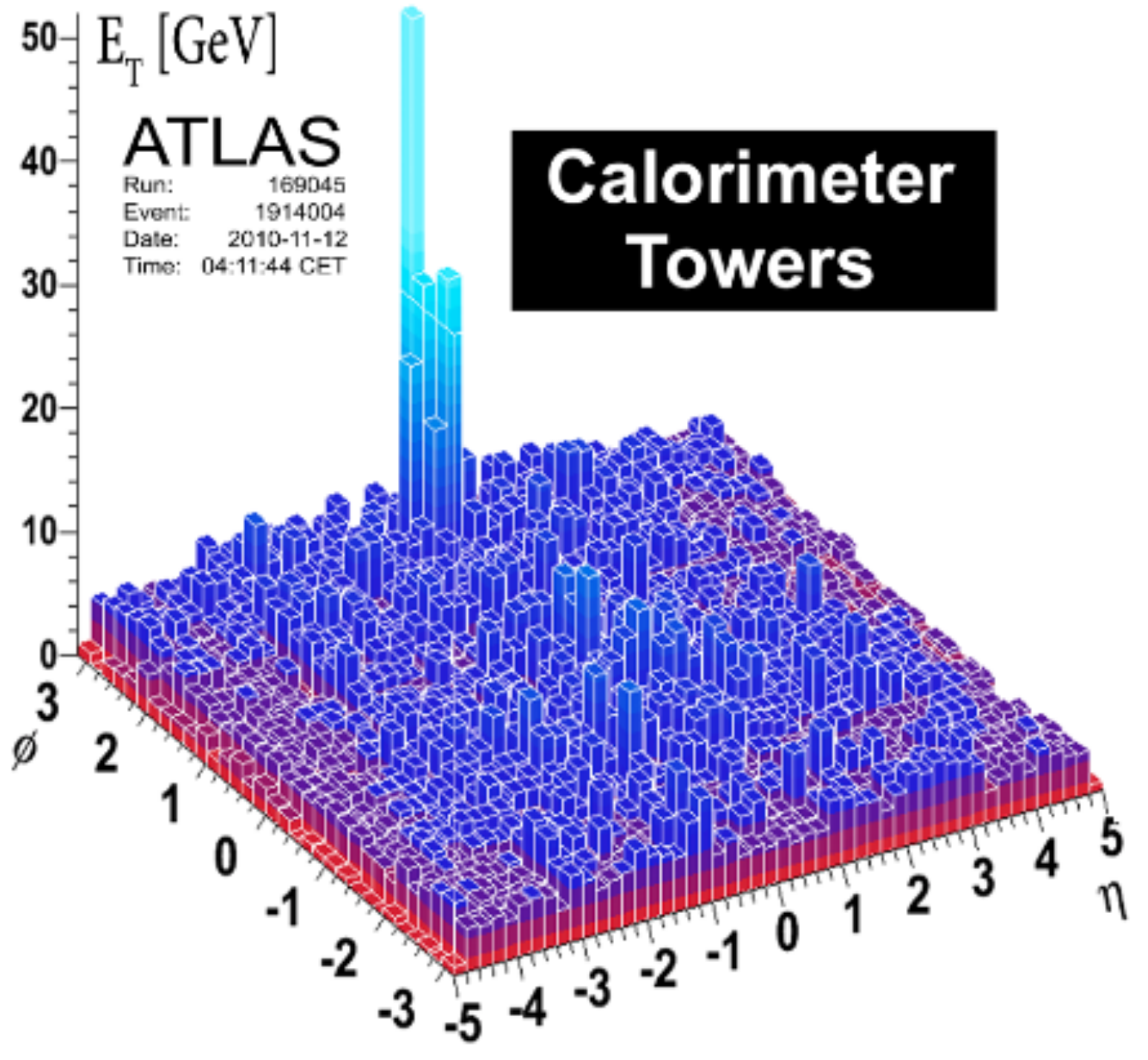}
\includegraphics[width=0.46\columnwidth]{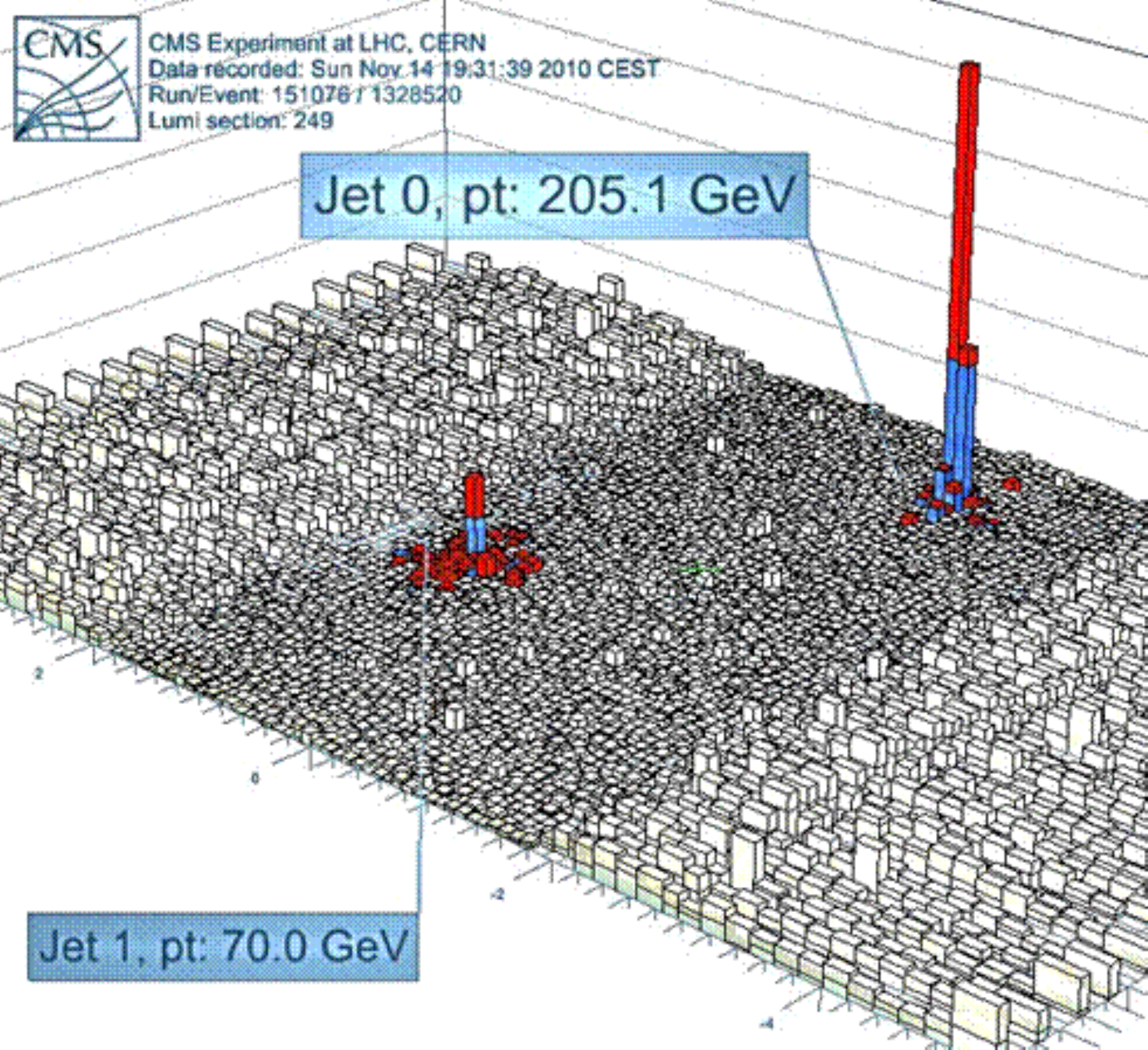}
\caption{Examples of unbalanced dijet events in PbPb collisions 
at $\sqrt{s_{NN}}= 2.76$~TeV. Figures are taken from ~\cite{ATLAS_jet} and ~\cite{CMS_jet}.
}
\label{fig:jet}
\end{center}
\end{figure}

\subsection{$R_{AA}$ of Charged particle and photons}

The nuclear modification factor, $R_{\rm AA}$, which gives the 
modification of the charged particle \pt spectrum, compared to
nucleon--nucleon collisions at the same energy, can shed light on the
detailed mechanism by which hard partons lose energy traversing the
medium. 
$R_{\rm AA}$ has been measured in CMS for all charged particles with
\pt up to 100 GeV/c. 
The ratio is plotted in the left panel of Fig.~\ref{fig:Raa} 
and compared to theoretical predictions. A stronger suppression
compared to those of the RHIC data has been observed.

Direct photon production as a function of centrality and \pt has been
made by the CMS experiment~\cite{CMS_mult}. The ratio, $R_{\rm AA}$ as
a function of photon \pt is shown in the right panel of
Fig.~\ref{fig:Raa}. The results are compared to the NLO pQCD
predictions. Within statistical uncertainties, the direct photons are
seen to be not suppressed compared to \mbox{p-p} collisions.

\subsection{Jet quenching and dijet asymmetry}

\begin{figure}[ht]
\begin{center}
\includegraphics[width=0.9\columnwidth]{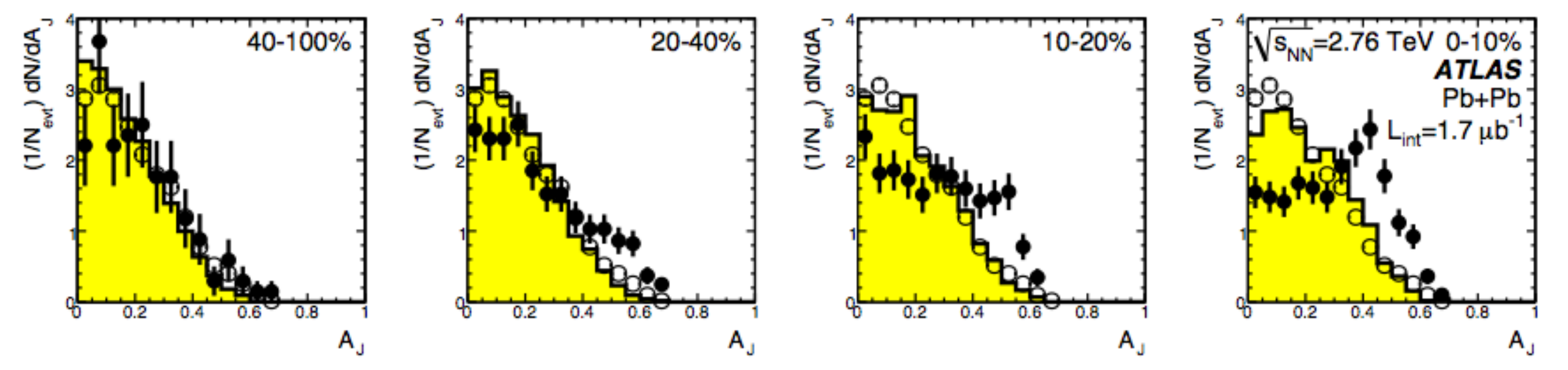}
\includegraphics[width=0.7\columnwidth]{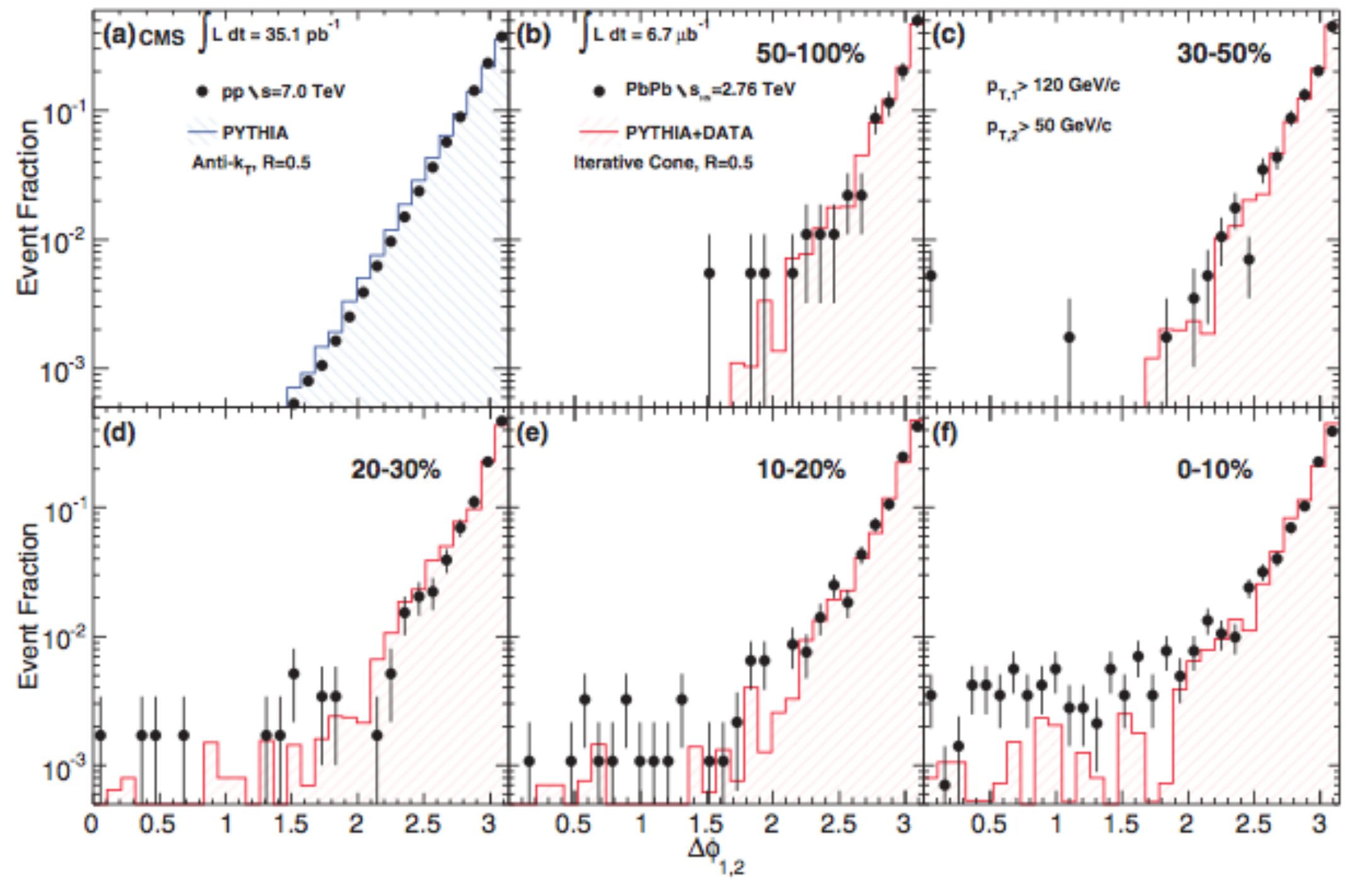}
\caption{(Top) Dijet asymmetry for different centrality bins for
\mbox{Pb-Pb} collisions, showing a strong centrality dependence. 
(Below) Azimuthal angle distributions between dijets for different
centrality bins~\cite{ATLAS_jet} and ~\cite{CMS_jet}.
}
\label{fig:dijet}
\end{center}
\end{figure}

At the LHC, very high \pt jets are most of the time easily visible
above the soft background~\cite{ATLAS_jet,CMS_jet} because of the
availability of large energy. The medium effects are studied by
analyzing the transverse energy of dijets in opposite hemispheres. 
Fig.~\ref{fig:jet} shows typical events in both ATLAS and CMS
experiments with large imbalance in energy from one of the hemisphere
to the other. This is a very striking observation and direct evidence
of jet quenching.
The transverse energies of dijets in opposite
hemispheres is observed to become systematically more unbalanced with
increasing event centrality leading to a large number of events which
contain highly asymmetric dijets. One of the main question is to
find out what happened to the energy in the opposite hemisphere.

Quantitative measurements of dijet asymmetry have been reported by 
both ATLAS~\cite{ATLAS_jet}  and  CMS~\cite{CMS_jet} experiments. 
The upper panel of Fig.~\ref{fig:dijet} shows the dijet asymmetry for
peripheral to central collisions. 
Proton-proton
data from	$\sqrt s = 7$~TeV, 
analyzed with the same jet selection, is shown as open circles,
whereas the solid histograms depict the results from PYTHIA dijet simulations.
For \mbox{p-p} and for peripheral
collisions, the asymmetry value is peaked at or close
to zero, implying the jets are balanced on both hemispheres. 
For central collisions, a marked asymmetry develops with unbalanced
jets ($A_j>0$). This shows an enhancement of events with large dijet
asymmetries, not observed in \mbox{p-p} collisions.
The bottom panels of Fig.~\ref{fig:dijet} shows the CMS results on the
azimuthal angle between the dijets for \mbox{p-p} and for
different centrality ranges in \mbox{Pb-Pb}.
In all cases, the two
jets stay mostly back-to-back in azimuth. 
CMS has also measured the distribution of energy around the jets with
most of the lost energy appearing in very low momentum fragments far 
away from the jet direction~\cite{CMS_jet}. 
Results from both ATLAS and CMS point to jet quenching and an interesting 
new mechanism where the energy is
radiated away via multiple, soft gluons, which in turn may even
re-interact in the medium and lead to a further degradation of the
energy.

\section{Summary and outlook}

With the acceleration of heavy ions in the year 2010, the LHC has
entered a new domain of physics. The first data, recorded and analyzed by ALICE,
ATLAS and CMS experiments have been most impressive in terms of the
results, some of which are consistent with present understandings and
some others which are not. The salient features of the
latest findings are as follows:

\begin{itemize}
\item The results of the multiplicity distributions, fluctuations and 
Bose-Einstein correlations point to the formation of a fireball 
which is  hotter, long lived and larger in size, much more than
those measured at RHIC.

\item Measurements of anisotropic flow infer to the formation of a system
which behaves like an almost perfect fluid with almost no friction.
This will provide strong constraints on the temperature dependence of
$\eta/s$. 

\item The $J/\psi$ suppression is found to be similar in
magnitude to SPS and RHIC. One of the possible explanations of this
observation is that the suppression of $J/\psi$  may be
getting balanced by coalescence of two independently created charm quarks. 
Detailed measurement of both $J/\psi$ and $\Upsilon$ families will be needed to answer this.

\item In central collisions, a large suppression of heavy flavours is
observed,  indicating that  charm quarks undergo a strong energy loss
in the hot and dense matter.

\item Electroweak $W$ and $Z$ boson measurements have become possible for the first time
in heavy ion collisions. These channels may indeed provide most
sensitive probes of the initial state of QGP matter.

\item Unexpectedly large dijet asymmetry and strong jet energy loss
  have been observed, suggesting strong interactions between jets and a hot, dense medium.

\end{itemize}

The available 
results take us one step closer to understanding the
primordial state of matter that existed within few microseconds after the Big Bang.
Better understanding is expected in near future with
more exclusive measurements and more theoretical studies.
%In near future, LHC is expected to run at higher luminosity and also collide
%\mbox{p-Pb}, both of which will be quite valuable.

\bigskip
\noindent
{\bf Acknowledgement:} We would like to thank the Organizers of Lepton-Photon 2011 for the
invitation to present LHC heavy ion results.
We would like to thank the LHC Operations for providing
excellent beams, and to the members of ALICE,
ATLAS and CMS Collaborations for the analysis results. 
In particular, we would like to express our thanks to Panos
Christakoglou, Premomoy Ghosh, Paolo Giubellino, Satyajit Jena, Sanjib
Muhuri, Peter Steinberg,
Ermano Vercelli,  and Bolek Wyslouch
for their help during preparation of the talk and the manuscript.

\bibliographystyle{pramana}
\bibliography{references}

\begin{thebibliography}{99}

\bibitem{satz}         H.~Satz, Nucl. Phys. {\bf A862}, (2011) 4.

\bibitem{kapusta}      J.~Kapusta, Nucl. Phys. {\bf A862}, (2011) 47.

\bibitem{ALICE}     K.~Aamodt {\em et al.} (ALICE Collaboration),
  JINST {\bf 3}, (2008) S08002.

\bibitem{ATLAS}     G. Aad  {\em et al.} (ATLAS Collaboration), JINST {\bf 3}, (2008) S08003.

\bibitem{CMS}  R.~Adolphi {\em et al.} (CMS Collaboration), JINST
{\bf 3}, (2008) S08004.

%ALICE centrality

\bibitem{ALICEcharged} K.~Aamodt {\em et. al.} (ALICE Collaboration)
  Phys. Rev. Lett. {\bf 106}, (2011) 032301.

% Multiplicity distribution

\bibitem{ALICE_mult} K. Aamodt {\it et al.} (ALICE Collaboration)
  Phys. Rev. Lett. {\bf 105} (2010) 252301 (2010). 
\bibitem{CMS_mult} B. Wyslouch {\em et al.} (CMS Collaboration)
J. Phys. G: Nucl. Part. Phys. {\bf 38} (2011) 124005.
\bibitem{ATLAS_mult}  G. Aad {\em et al.}  (ATLAS Collaboration), [arXiv:1108.6027 [hep-ex]].

% HBT

\bibitem{ALICE_hbt} K. Aamodt {\em et al.} (ALICE Collaboration)
Phys. Lett. {\bf B696} (2011) 328.
\bibitem{debasish} B.I. Abelev {\it et al.} (STAR Collaboration)
Phys. Rev. {\bf C80} (2009) 024905.

% Flow
\bibitem{ALICE_flow} K. Aamodt {\em et al.} (ALICE Collaboration) Phys. Rev. Lett. {\bf
105} (2010) 252302.
\bibitem{raimond} R. Snellings {\em et al.} (ALICE Collaboration) J. Phys. G:
  Nucl. Part. Phys. {\bf 38} (2011) 124013.
\bibitem{jurgen} J. Schukraft, arXiv:1112.0550v1 [hep-ex].
\bibitem{kovtun} P. Kovtun, D. T. Son, A. O. Starinets,
Phys. Rev. Lett. {\bf94} (2005) 111601.

% Fluctuation

\bibitem{JeonKoch00}   S.~Jeon, V.~Koch, Phys. Rev. Lett. {\bf 85}, (2000) 2076.
\bibitem{pruneau} C. Pruneau, S. Gavin and S. Voloshin,
 Phys. Rev. {\bf C 66} (2002) 044904.
\bibitem{StarChargeFluctuations} B.~I.~Abelev {\em et al.}
    (STAR Collaboration), Phys. Rev. {\bf C79}, (2009) 024906.

\bibitem{panos} P. Christakoglou {\em et al.} (ALICE Collaboration), arXiv:1111.4506v1 [nucl-ex].

\bibitem{satya} S. Jena {\em et al.} (ALICE  Collaboration), arXiv:1201.0130 [hep-ex].

% Quarkonia
\bibitem{ALICE_quarkonia1} P. Pillot {\em et al.} ) (ALICE Collaboration) J. Phys.  {\bf G38} (2011) 124111.
\bibitem{ALICE_quarkonia2} D. Das {\em et al.} ) (ALICE
 Collaboration) arXiv:1111.5946v1  [nucl-ex].
\bibitem{CMS_quarkonia} S. Chatrchyan {\em et al.} (CMS
  Collaboration), Phys. Rev. Lett. {\bf 107} (2011) 052302.


% Heavy Flavor
\bibitem{ALICE_dainese} A. Dainese {\em et al.} (ALICE Collaboration)
    J. Phys. G: Nucl. Part. Phys. {\bf 38} (2011) 124032.

% Electroweak

\bibitem{ATLAS_weak} R Sandström {\em et al.} (ATLAS Collaboration)
J. Phys. G: Nucl. Part. Phys. {\bf 38} (2011)  124133.
\bibitem{CMS_weak} S. Chatrchyan {\em et al.} (CMS
 Collaboration), Phys. Rev. Lett. {\bf 106} (2011) 212301.

% energy loss


%  Jet quenching
\bibitem{ATLAS_jet} G. Aad {\em et al.}  (ATLAS Collaboration)
Phys. Rev. Lett. {\bf 105} (2010) 252303.
\bibitem{CMS_jet} S. Chatrchyan {\em et al.} (CMS Collaboration)
  Phys. Rev. {\bf C84} (2011) 024906.

 \end{thebibliography}

%\bibitem{ATLASpeter} P. Steinberg in the Proceedinngs of the DPF-2011
 % Conference, Providence, RI, Aug 8-13, 2011, arXiv:1110.3352v1 [nucl-ex]

% \begin{thebibliography}{99}
% \bibitem{lever}
% Author Name, {\it Book Name} (Publisher, Address, 1998) 
% 
% \bibitem{otherref}
% Some other, {\it Name of some journal}, {\bf 1234}, 56, (1202)
% 
% \bibitem{kinginexile} 
% This and that author, {\it Name of journal}, {\bf 3904}, 230, (2009).
% This is an additional note of no importance.
% \end{thebibliography}

\end{document}